\documentclass[a4paper,11pt, reqno]{amsart}
\usepackage{eufrak}
\usepackage[all]{xy} 
\usepackage{amssymb} 
\usepackage{epsfig}
\usepackage{multicol}
\usepackage{xypic}
\usepackage{amsmath}
\usepackage{wasysym}
\newcommand{\lp}{\left(}
\newcommand{\rp}{\right)}
\newcommand{\lc}{\left\{}
\newcommand{\rc}{\right\}}
\newcommand{\der}{\partial}
\newcommand{\cua}{^{2}}

\newcommand{\bra}{\langle}
\newcommand{\ket}{\rangle}
\newcommand{\R}{\mathbb{R}}      
\newcommand{\F}{\mathbb{F}}
\newcommand{\LL}{\mathbb{L}}
\newcommand{\Flder}{\rightarrow}
\newcommand{\qmid}{\frac{q_{1}+q_{0}}{2}}
\newcommand{\pmid}{\frac{p_{1}+p_{0}}{2}}
\newcommand{\Ups}{\Upsilon^{-1}(\Lambda)}
\newcommand{\pmed}{p_{k+1/2}}
\newcommand{\qmed}{q_{k+1/2}}
\newcommand{\qk}{q_{k}}
\newcommand{\pk}{p_{k}}
\newcommand{\qkm}{q_{k+1}}
\newcommand{\pkm}{p_{k+1}}
\newcommand{\de}{\mbox{d}}

\newcommand{\be}{\begin{equation}}
\newcommand{\ee}{\end{equation}}
\newtheorem{definition}{Definition}[section]

\newtheorem{theorem}[definition]{Theorem}

\newtheorem{remarkth}[definition]{Remark}

\newenvironment{remark}{\begin{remarkth}\upshape}{\end{remarkth}}

\begin{document}

\title[Hamiltonian dynamics and constrained variational calculus]{Hamiltonian dynamics and constrained variational calculus: continuous and discrete settings}
\author[M. de Le\'on]{Manuel de Le\'on}
\address{M. de Le\'on: Instituto de Ciencias Matem\'aticas (CSIC-UAM-UC3M-UCM), Serrano 123, 28006
Madrid, Spain} \email{mdeleon@icmat.es}

\author[F. Jim\'enez]{Fernando Jim\'enez}
\address{F. Jim\'enez: Instituto de Ciencias Matem\'aticas (CSIC-UAM-UC3M-UCM), Serrano 123, 28006
Madrid, Spain} \email{fernando.jimenez@icmat.es}

\author[D.\ Mart\'{\i}n de Diego]{David Mart\'{\i}n de Diego}
\address{D.\ Mart\'{\i}n de Diego: Instituto de Ciencias Matem\'aticas (CSIC-UAM-UC3M-UCM), Serrano 123, 28006
Madrid, Spain} \email{david.martin@icmat.es}

\maketitle

\begin{abstract}
The aim of this paper is to study the relationship between Hamiltonian dynamics and constrained variational calculus. We describe both using the notion of Lagrangian submanifolds of convenient
symplectic manifolds and using the so-called Tulczyjew's triples. The results are also extended to the case
of discrete dynamics and nonholonomic mechanics. Interesting applications to geometrical integration of Hamiltonian systems are obtained.
\end{abstract}

\section{Introduction}

One of the main notions  in symplectic geometry is the concept of  Lagrangian submanifolds. This concept arises in several and different interpretations of physical, engineering and  geometric phenomena. In this paper, we will focus our attention in their applications to Lagrangian and Hamiltonian dynamics of constrained systems.

For instance, the theory of Lagrangian submanifolds gives a geometric and intrinsic description of Lagrangian and Hamiltonian dynamics \cite{Tulczy1,Tulczy2}.  Moreover, it allows us to relate both formalisms using as a main tool the so-called Tulczyjew's triple
\[
\xymatrix{
T^{*}TM&TT^{*}M\ar[l]_{\alpha_{M}}\ar[r]^{\beta_{M}}&T^{*}T^*M.
}
\]
Recall that, in the above expression, $\alpha_{M}$ is  the Tulczyjew's canonical symplectomorphism from $TT^{*}M$ (with
its canonical symplectic structure \linebreak $\mathrm{d}_T{\omega}_M$) to $T^*TM$ (equipped now with its canonical symplectic structure $\omega_{TM}$);  and
$\beta_{M}$ is the canonical symplectomorphism defined by the symplectic structure $\omega_M$ on $T^*M$.
The Lagrangian dynamics is ``generated'' by the Lagrangian submanifold $\mathrm{d}\LL(TM)$  of $T^*TM$ where $\LL: TM\rightarrow \R$ is the Lagrangian function, while the Hamiltonian formalism is generated by the Lagrangian submanifold $\mathrm{d}H(T^*M)$  of $T^*T^*M$ where $H:T^*M\rightarrow \R$ is the corresponding Hamiltonian energy. The dynamics and the relationship between both formalisms are based on the central part of the
Tulczyjew's triple, $TT^*M$, where the Lagrangian submanifolds $\alpha_M(\mathrm{d}\LL(TM))$ and  $\beta^{-1}_M(\mathrm{d}H(T^*M))$ live. Of course, any submanifold of a tangent bundle automatically determines a system of implicit differential equations; in this case, we can apply the integrability constraint algorithm, described in $\S$ \ref{ALG} (see \cite{CLMM,Mendella} for more details), to find the integrable part of the dynamics defined by this Lagrangian submanifold.

This model is also valid for constrained variational calculus, determined by a function $L: C\to \R$ where $C$ is a submanifold of $TM$, with inclusion $i_C: C\to TM$. In this case, we can also construct a new Lagrangian submanifold (see $\S$ \ref{sigma}) $\Sigma_{L}$ of $T^{*}TM$ as:
\[
\Sigma_{L}=\{\mu\in T^*TM\, |\; i_C^*\,\mu=\mathrm{d}L\}.
\]
Thus, we can obtain via $\alpha_{M}$ a new Lagrangian submanifold of the tangent bundle $TT^*M$ which completely determines the equations of motion of the constrained dynamics (see \cite{GG}), which are, in a regular case, of Hamiltonian type. In this paper, we will also prove that given an arbitrary Hamiltonian system we can construct a (possibly) constrained Lagrangian system that generates the original one.

It is necessary to stress that the equations derived are purely variational and, consequently, different from the nonholonomic equations obtained by applying the Lagrange-D'Alembert's or Chetaev's principles (see \cite{Bl,Cort,Manolo} for more details). It is well-known that the nonholonomic equations give the  right physical dynamics of a constrained mechanical systems \cite{lm}, mainly related to the rolling motion. Geometrically, nonholonomic constraints are globally described by
a submanifold $C$ of the velocity phase space $TM$. If
$C$ is a vector subbundle of $TM$, we are dealing with
linear constraints and, when $C$ is an
affine subbundle, we are in the case of affine constraints.
Lagrange-D'Alembert's  or Chetaev's principles allow us to
determine the set of possible values of the constraint forces only
from the set of admissible kinematic states, that is, from the
constraint submanifold $C$ determined by the vanishing of
the nonholonomic constraints. An interesting study of nonholonomic systems as implicit differential equations is presented in \cite{Italia}.

On the other hand, constrained variational calculus is mainly related to mathematical and engineering applications, specially in control theory. In this paper, we will show the close relationship between classical Hamiltonian dynamics and constrained variational calculus. In fact, both are equivalent under some regularity conditions. Nevertheless, nonholonomic mechanics is described by a general submanifold of $TT^*M$ which is not Lagrangian; this fact implies the non-preservation properties of the nonholonomic flow.

Moreover, we will study the discrete formalism, which will be also interpreted in the same way as  Lagrangian submanifolds of the cartesian product of two copies of $T^*M$, equipped  with a suitable symplectic structure. We are interested in finding a geometrical answer to two, in principle, alternative ways to derive geometric integrators for constrained Lagrangian problems: a pure discrete variational procedure or a symplectic numerical method for the associated Hamiltonian system. We will show in which cases both procedures match. Additionally, we will derive a general algorithm to compare the dynamics in variational constrained and nonholonomic cases, both in continuous and discrete cases.


\section{Geometric preliminaries}

\subsection{Lagrangian submanifolds}
In this section we will introduce some particular constructions of Lagrangian submanifolds that are interesting for our purposes (see \cite{LiMa,Wein}).

First, let us recall that given a finite-dimensional symplectic manifold $(P, \omega)$ and a submanifold $N$, with canonical inclusion $i_N: N\hookrightarrow P$, then $N$ is a \emph{Lagrangian submanifold}  if
 $i^{*}_{N}\,\omega=0$ and $\mbox{dim}\hspace{1mm}N=\frac{1}{2}\mbox{dim}\hspace{1mm}P$.

\begin{itemize}
\item[$i)$] An interesting class of Lagrangian submanifolds, which will be useful in $\S$ \ref{DEQ}, is the following.
Let $\lp P,\omega\rp$ be a symplectic manifold and $g:P\Flder P$ a diffeomorphism. Denote by $\mbox{Graph}\lp g\rp$ the graph of $g$, that is $\mbox{Graph}\lp g\rp=\lc\lp x,g\lp x\rp\rp,\hspace{1mm} x\in P \rc\subset P\times P$, and by $pr_{i}:P\times P\Flder P$, $i=\lc 0,1\rc$, the canonical projections. Then  $\lp P\times P,\Omega\rp$, where $\Omega=pr^{*}_{1}\hspace{1mm}\omega-pr^{*}_{0}\hspace{1mm}\omega$, is a symplectic manifold. Let $i_{g}:\mbox{Graph}\lp g\rp\hookrightarrow P\times P$ be the inclusion map, then
$$
i_{g}^{*}\Omega=\lp pr_{0}\rp^{*}\lp g^{*}\omega-\omega\rp.
$$
Thus, $g$ is a symplectomorphism (that is, $g^{*}\omega=\omega$) if and only if $\mbox{Graph}\hspace{1mm}g$ is a Lagrangian submanifold of $P\times P$.
\end{itemize}

A distinguished symplectic manifold is the cotangent bundle  $T^*M$ of any manifold $M$. If we choose local  coordinates  $(q^i)$, $1\leq i\leq n$, then $T^*M$ has induced  coordinates $(q^i, p_i)$. Denote by $\pi_M: T^*M\to M$ the canonical projection defined by $\pi_M(\epsilon_q)=q$, where $\epsilon_q\in T^*_{q}M$.
Define the Liouville one-form or canonical one-form $\theta_M\in \Lambda^1 T^*M$ by
\[
\bra(\theta_M)_{\epsilon}\,,\,X\ket=\bra\epsilon\,,\,T\pi_M(X)\ket, \hbox{  where  } X\in T_{\epsilon}T^*M\; ,\ \epsilon\in T^*M.
\]
In local coordinates we obtain $\theta_M=p_i\, \mathrm{d}q^i$.
The canonical two-form $\omega_M$ on $T^*M$ is the symplectic form $\omega_M=-\mathrm{d}\theta_M$ (that is $\omega_{M}=\mathrm{d}q^{i}\wedge\mathrm{d}p_{i}$).
\begin{itemize}
\item[$ii)$] Now, we will introduce some special Lagrangian submanifolds of the symplectic manifold $(T^*M, \omega_M)$.
For instance, the image ${\Sigma}_{\lambda}=\lambda(M)\subset T^*M$ of a closed one-form $\lambda\in \Lambda^1M$ is a Lagrangian submanifold of $(T^*M, \omega_M)$,
since $\lambda^* \omega_M = - \mathrm{d}\lambda$. We then obtain a submanifold diffeomorphic to $M$ and transverse to the fibers of $T^*M$. When $\lambda$ is exact, that is, $\lambda=\mathrm{d}f$, where $f: M\to \R$, we say that $f$ is a \emph{generating} function of the Lagrangian submanifold ${\Sigma}_{\lambda}=\Sigma_{f}$. Locally, this is always the case.
\end{itemize}
A useful extension of the previous construction is the following result due to W.M. Tulczyjew.

\begin{theorem}[\cite{Tulczy1},\cite{Tulczy2}]\label{tulchi}
  \label{thm:tulczyjew}
  Let $M$ be a smooth manifold, $\tau_M:TM\Flder M$ its tangent bundle projection, $ N \subset M $ a submanifold, and $ f
  \colon N \rightarrow \mathbb{R} $.  Then
  \begin{multline*}
    \Sigma _f = \bigl\{ p \in T ^\ast M \mid \pi _M (p) \in N \text{
        and } \left\langle p, v \right\rangle = \left\langle
        \mathrm{d} f , v \right\rangle \\
      \text{ for all } v \in T N \subset T M \text{ such that } \tau
      _M (v) = \pi _M (p) \bigr\}
  \end{multline*}
  is a Lagrangian submanifold of $ T ^\ast M $.
\end{theorem}
Taking $f$ as the zero function we obtain the following Lagrangian submanifold
\[
\Sigma_{0}=\lc p\in T^{*}M\big|_{N}\,|\,\bra p\,,\,v\ket=0\,,\,\forall\,v\in TN\,\mbox{with}\,\tau_{M}(v)=\pi_{M}(p)\rc,
\]
which is just the {\bf conormal bundle} of $N$:
\[
\nu^{*}(N)=\lc\xi\in T^{*}M\big|_{N}\,;\,\xi\big|_{T_{\pi(\xi)}N}=0\rc.
\]

Given  a symplectic manifold $(P, \omega)$, $\dim P=2n$ it is well-known that its tangent bundle $TP$ is equipped with a
symplectic structure denoted by $\mathrm{d}_T\omega$ (see \cite{dlr}).
If we take Darboux coordinates $(q^i,p_i)$ on $P$, $1\leq i\leq n$, then $\omega=\mathrm{d}q^i\wedge \mathrm{d}p_i$ and, consequently, we have induced coordinates $(q^i, p_i; \dot{q}^i, \dot{p}_i)$, $(q^i, p_i; a_i, b^i)$ on $TP$ and $T^*P$, respectively. Thus,
$\mathrm{d}_T\omega=\mathrm{d}\dot{q}^i\wedge \mathrm{d}p_i+\mathrm{d}q^i\wedge \mathrm{d}\dot{p}_i$ and $\omega_{P}=\mathrm{d}q^i\wedge \mathrm{d}a_i+\mathrm{d}p_i\wedge \mathrm{d}b^i$.
If we denote by $\flat_{\omega}: TP\to T^*P$ the isomorphism defined by $\omega$, that is $\flat_{\omega}(v)=i_{v}\,\omega$, then
we have $\flat_{\omega}(q^i, p_i; \dot{q}^i, \dot{p}_i)=(q^i, p_i; -\dot{p}_i, \dot{q}^i)$.
Given a function $H: P\to \R$, and its associated Hamiltonian vector field $X_H$, that is, $i_{X_H}\omega_{P}=\mathrm{d}H$, the image $X_H(P)$ is a Lagrangian submanifold of  $(TP, \mathrm{d}_T\omega_P)$. Moreover, given a vector field $X\in {\mathfrak X}(P)$, it is locally Hamiltonian if and only if its image $X(P)$ is a Lagrangian submanifold of $(TP, \mathrm{d}_T\omega)$.
It is interesting to note that
$\mathrm{d}_T\omega=-\flat_{\omega}^{*}\, \omega_M$ and $\flat_{\omega}(X_H(M))=\mathrm{d}H(M)$.

As it is briefly mentioned above, an important notion in the theory of Lagrangian submanifolds is the concept of generating function. If we have a Lagrangian submanifold $N$ of an exact symplectic manifold $(P,\omega=\mathrm{d}\theta)$, where $\theta\in \Lambda^1P$,  then
$0=i_N^*\omega=i_N^*\mathrm{d}\theta=\mathrm{d}(i_N^*\theta)$. Consequently, applying the Poincar\'e's lemma,  there exists a function $S: U\to \R$ defined on a open neihborhood $U$ of $N$ such  that $i_N^*\theta=\mathrm{d}S$. We say that $S$ is a (local) \emph{generating function} of the Lagrangian submanifold $N$.

\subsection{Implicit differential equations}\label{ALG}
An implicit differential equation on a general smooth manifold $M$ is a submanifold $E\subset TM$. A solution of $E$ is any curve $\gamma: I\Flder M$, $I\subset\R$, such that the tangent curve $(\gamma(t),\dot\gamma(t))\in E$ for all $t\in I$. The implicit differential equation will be said to be {\it integrable at a point} if there exists a solution $\gamma$ of $E$ such that the tangent curve passes through it. Furthermore, the implicit differential equation will be said to be {\it integrable} if it is integrable at {\it all points}. Unfortunately, integrability does not mean uniqueness. The integrable part of $E$ is the subset of all integrable points of $E$. The {\it integrability problem} consists in identifying such a subset.

Denoting the canonical projection $\tau_M:TM\Flder M$, a sufficient condition for the integrability of $E$ is
\[
E\subset TM,
\]
where $C=\tau_M(E)$, provided that the projection $\tau_M$ restricted is a submersion onto $C$.
\subsubsection{Extracting the integrable part of $E$} A recursive algorithm was presented in \cite{Mendella} that allows to extract the integrable part of an implicit differential equation $E$. We shall define the subsets
\[
E_0=E,\,\,\,\,C_0=C,
\]
and recursively for every $k\geq1$,
\[
E_k=E_{k-1}\cap TC_{k-1},\,\,\,\,C_k=\tau_M(E_k),
\]
then, eventually the recursive construction will stabilize in the sense that $E_k=E_{k+1}=...=E_{\infty}$, and $C_k=C_{k+1}=...=C_{\infty}$. It is clear by construction that $E_{\infty}\subset TC_{\infty}$. Then, provided that the adequate regularity conditions are satisfied during the application of the algorithm, the implicit differential equations $E_{\infty}$ will be integrable and it will solve the integrability problem

\section{Tulczyjew's triples}

In this section we summarize a classical result due to W.M. Tulczyjew showing a natural identification of $T^*TM$ and $TT^*M$, where $M$ is any smooth manifold, as symplectic manifolds. This construction plays a key role in Lagrangian and Hamiltonian mechanics.

Is easy to see that $TT^{*}M$, $T^{*}TM$ and $T^{*}T^{*}M$ are naturally {\it double vector bundles} (see \cite{Godbillon}, \cite{Pradines}) over $T^{*}M$ and $TM$. In \cite{Tulczy1} and \cite{Tulczy2}, Tulczyjew established two identifications, the first  one between $TT^{*}M$ and $T^{*}TM$ (useful to describe Lagrangian mechanics) and the second one between $TT^{*}M$ and $T^{*}T^{*}M$ (useful to describe Hamiltonian mechanics). The Tulczyjew map $\alpha_{M}$ is an isomorphism between $TT^*M$ and $T^*TM$. Beside, it is also a symplectomorphism between these double vector bundles  as symplectic manifolds, i.e. $(TT^*M\,,\,\mathrm{d}_T\,\omega_M)$, where $\mathrm{d}_T\,\omega_M$ is the tangent lift of $\omega_M$, and $(T^{*}TM, \omega_{TM})$. In the following diagram we show the different relationships among these bundles.
\[
\xymatrix{
&TT^{*}M\ar[rr]^{\alpha_{M}}\ar[ddl]_{\tau_{T^{*}M}}\ar[ddr]^{T\pi_{M}}& &T^{*}TM\ar[ddl]_{\pi_{TM}}\ar[ddr]^{T^{*}\tau_{M}}&\\
\\
T^{*}M\ar[ddrr]_{\pi_{M}}& &TM\ar[dd]_{\tau_{M}}& &T^{*}M\ar[ddll]^{\pi_{M}}\\\\
& & M& &
}
\]

The definition of $T^{*}\tau_{M}$ is given in the following remark.

\begin{remark}
Given a tangent bundle $\tau_{N}:TN\Flder N$, for each $y\in T_{x}N$ we can define
\[
{\mathcal V}_{y}=\mbox{ker}\lc T_{y}\tau_{N}:T_{y}TN\Flder T_{x}N\rc,\,\,\,\,\,\tau_{N}(y)=x.
\]
Summing over all $y$ we obtain a vector bundle ${\mathcal V}$ of rank $n$ over $TN$.
Any element $u\in T_{x}N$  determines a vertical
vector  at  any point  $y$ in the  fibre  over $x$, called its vertical lift to  $y$,  denoted
by $u^V(y)$. It is the tangent vector
at $t=0$ to the curve $y+t\,u$.
If $X$ is a vector field on $N$, we may define its vertical lift as $X^{V}(y)=\lp X(\tau_{N}(y))\rp^{V}$. Locally, if $X=X^{i}\frac{\der}{\der x^{i}}$ in a local neighborhood $U$ with local coordinates $x^{i}$, then $X^{V}$ is locally given by
\[
X^{V}=X^{i}\frac{\der}{\der v^{i}},
\]
with respect to induced coordinates $(x^{i},v^{i})$ on $TU$.

 Now, we define $T^*\tau_{M}: T^*TM\to T^*M$ by $\langle T^*\tau_{M} ( \alpha_{u}), w\rangle=\langle \alpha_u, w^V_{u}\rangle$; $u, w\in T_qM$, $\alpha_u\in T^*_uTM$  and $w^V_u\in T_u TM$.

\end{remark}

In the following, we recall the construction of the symplectomorphism $\alpha_M$. To do this, it is necessary to introduce the canonical flip (\cite{Godbillon}) on $TTM$:
\begin{equation*}
\xymatrix{
TTM\ar[d]_{\tau_{TM}}\ar[r]_{\kappa_{M}}&TTM\ar[d]^{T\tau_{M}}\\
TM\ar[r]_{\mbox{Id}}&TM,
}
\end{equation*}
 as follows:
$$
\kappa_{M}\lp\frac{\mathrm{d}}{\mathrm{d}s}\Big |_{s=0}\frac{\mathrm{d}}{\mathrm{d}t}\Big |_{t=0}\hspace{1mm}\chi\lp s,t\rp\rp=\frac{\mathrm{d}}{\mathrm{d}s}\Big |_{s=0}\frac{\mathrm{d}}{\mathrm{d}t}\Big |_{t=0}\hspace{1mm}\tilde\chi\lp s,t\rp,
$$
where $\chi:\R\cua\rightarrow M$ and $\tilde\chi:\R\cua\rightarrow M$ are related by $\tilde\chi\lp s,t\rp=\chi\lp t,s\rp$. If $\lp q^{i}\rp$ are the local coordinates for $M$, $\lp q^{i},v^{i}\rp$ for $TM$ and $\lp q^{i},v^{i},\dot q^{i},\dot v^{i} \rp$ for $TTM$, then the canonical involution can be defined as $\kappa_{M}\lp q^{i},v^{i},\dot q^{i},\dot v^{i}\rp=\lp q^{i},\dot q^{i},v^{i},\dot v^{i}\rp$.

In order to describe $\alpha_{M}$ is also necessary to define a tangent pairing. Given two manifolds $M$ and $N$, and a pairing between them $\bra\cdot,\cdot\ket:M\times N\rightarrow\R$, the tangent pairing $\bra\cdot,\cdot\ket^{T}:TM\times TN\rightarrow\R$ is determined by
$$
\bra\frac{\mathrm{d}}{\mathrm{d}t}\Big |_{t=0}\hspace{1mm}\gamma\lp t\rp,\frac{\mathrm{d}}{\mathrm{d}t}\Big |_{t=0}\hspace{1mm}\delta\lp t\rp\ket^{T}=\frac{\mathrm{d}}{\mathrm{d}t}\Big |_{t=0}\hspace{1mm}\bra\gamma\lp t\rp,\delta\lp t\rp\ket
$$
where $\gamma:\R\rightarrow M$ and $\delta:\R\rightarrow N$.

Finally, we can define  $\alpha_{M}$ as $\bra\alpha_{M}\lp z\rp,w\ket=\bra z,\kappa_{M}\lp w\rp\ket^{T}$, where $z\in TT^{*}M$ and $w\in TTM$. In local coordinates:
$$
\alpha_{M}\lp q^{i},p_{i},\dot q^{i},\dot p_{i}\rp=\lp q^{i},\dot q^{i},\dot p_{i},p_{i}\rp;
$$
now $\lp q^{i},p_{i}\rp$ are the local coordinates for $T^{*}M$ and $\lp q^{i},p_{i},\dot q^{i},\dot p_{i}\rp$ for $TT^{*}M$.

The third double vector bundle is $T^{*}T^{*}M$. The  isomorphism $\beta_{M}:TT^{*}M\rightarrow T^{*}T^{*}M$ is just given by $\beta_M=\flat_{\omega_M}$ (see previous subsection).

Considering the bundles $TT^{*}M$, $T^{*}TM$ and $T^{*}T^{*}M$, as well as the symplectomorphisms $\alpha_{M}$ and $\beta_{M}$, we finally obtain the Tulczyjew's triple
\begin{equation*}
\xymatrix{
T^{*}T^{*}M&TT^{*}M\ar[l]_{\beta_{M}}\ar[r]^{\alpha_{M}}&T^{*}TM\; .
}
\end{equation*}

\section{Continuous Lagrangian and Hamiltonian mechanics}
\label{sigma}

In the introduction we have shown how the Tulcyjew's triple is  used to describe geometrically Lagrangian and Hamiltonian mechanics and its relationship. In this section we will see that it is also possible to adapt this geometric formalism when we introduce constraints into the picture.
As we have mentioned along the introduction, there are (at least) two
methods that one might use to derive the equations of motion of systems subjected to constraints. We will call them nonholonomic mechanics and constrained variational calculus.
The classical method to derive equations of motion for constrained
mechanical systems is the nonholonomic mechanics. The equations derived from the nonholonomic methods {\it are not} of variational nature, but they describe the correct dynamics of a constrained mechanical system.
In order to obtain the nonholonomic equations, if we have linear or affine constraints, is necessary to apply the Lagrange-D'Alembert's principle. When dealing with nonlinear constraints, one should employ the more controversial Chetaev's rule (see \cite{Bl,Manolo} for further details). 	Since the geometrical implementation of the Chetaev's rule is practically equal to the process in the linear case, we shall use it from a pure mathematical perspective.

On the other hand, the equations of motion of constrained variational problems are derivable by
using variational techniques (always in the constrained case).
 These last equations are also known in the literature as vakonomic equations.  The terminology vakonomic (``mechanics of variational axiomatic kind'') was coined by V.V. Kozlov (\cite{Arnold},\cite{Koslov}). The main applications of the constrained variational calculus appear in problems of mathematical nature (like subriemannian geometry) and in optimal control theory.

\subsection{Nonholonomic mechanics}\label{NC}
A nonholonomic system on a manifold $M$ consists of a pair $(\LL, C)$, where $\LL: TM\rightarrow \R$ is the Lagrangian of the mechanical system and $C$ is a submanifold of $TM$ with canonical inclusion $i_C: C\hookrightarrow TM$. In the following, we will assume, for sake of symplicity, that $\tau_M(C)=M$. Since the motion of the system is forced to take place on the submanifold $C$, this requires the introduction of some reaction or constraint forces into the system.
If $\phi^{\alpha}(q^{i},\dot q^{i})=0$, $1\leq\alpha\leq n$, determine locally the submanifold $C$, then Chetaev's rule implies that the constrained equations of the system are:
\begin{eqnarray}\nonumber
&&\frac{\mathrm{d}}{\mathrm{d}t}\lp\frac{\der\LL}{\der\dot q^{i}}\rp-\frac{\der\LL}{\der q^{i}}=\lambda_{\alpha}\frac{\der\phi^{\alpha}}{\der\dot q^{i}},\\\label{Noholo}\\\nonumber
&&\phi^{\alpha}(q^{i},\dot q^{i})=0,\,\,\,\,1\leq\alpha\leq n.
\end{eqnarray}
Next, we will describe geometrically the nonholonomic equations. First, we need to introduce the vertical endomorphism $S$ which is a $(1,1)$-tensor field on $TM$ defined by
\[
\begin{array}{rrcl}
S:&TTM&\longrightarrow& TTM\\
  &W_{v_x}&\longmapsto&\displaystyle{\frac{\mathrm{d}}{\mathrm{d}t}\Big|_{t=0}}\lp v_x+t\,T\tau_M(W_{v_x})\rp.
\end{array}
\]
Its local expression is $S=\frac{\partial}{\partial\dot q^i}\otimes \mathrm{d}q^i$.

If we accept Chetaev-type forces, then we define
\[
F=S^*(TC)^{0}.
\]
Observe that the vector subbundle $F$ will be generated by the 1-forms $\mu^{\alpha}=\frac{\partial \phi^{\alpha}}{\partial \dot{q}^i}\; \mathrm{d}q^i$ because $S^*(\mathrm{d}\phi^{\alpha})=\mu^{\alpha}$.

Now, define the affine subbundle of $T^*_CTM$ given by
\[
\Sigma^{noh}=(\mathrm{d}\LL)\circ i_C+F,
 \]
that is,
\begin{eqnarray}
\Sigma^{noh}&=&\{
(q^i, \dot{q}^i, \mu_{i}, \tilde{\mu}_i)\in T^*TM\; |\;\label{sij}\\
&&\mu_i=\frac{\partial {\LL} }{\partial q^i} +\lambda_{\alpha}\frac{\partial \phi^{\alpha} }{\partial \dot{q}^i},\nonumber\\
&&\tilde{\mu}_i=\frac{\partial {\LL} }{\partial \dot{q}^i},\nonumber\\
&&\phi^{\alpha}(q,\dot q)=0,\,\,\, 1\leq\alpha\leq n\}\; .\nonumber
\end{eqnarray}
Therefore, applying the Tulczyjew's isomorphism $\alpha_M$ we obtain the affine subbundle
\begin{eqnarray}\label{Tulcnonho}
\alpha_M^{-1}\lp\Sigma^{noh}\rp&=&\{(q^i, p_i, \dot{q}^i, \dot{p}_i)\in TT^*M\;|\;\\
&&p_i=\frac{\partial {\LL} }{\partial \dot{q}^i} ,\nonumber\\\nonumber
&&\dot{p}_i=\frac{\partial {\LL} }{\partial q^i} +\lambda_{\alpha}\frac{\partial \phi^{\alpha} }{\partial \dot{q}^i},\\
&&\phi^{\alpha}(q,\dot q)=0,
\,\,\, 1\leq\alpha\leq n\}\; .\nonumber
\end{eqnarray}
Define now the nonholonomic  Legendre transformation $\F\LL^{noh}: C\rightarrow T^*M$ by
\[
\F\LL^{noh}=\pi_{T^*M}\circ \alpha_M^{-1}\circ d\LL\circ i_C\; .
\]
 The solutions for the dynamics given by $\alpha_M^{-1}\lp\Sigma^{noh}\rp$ are curves $\sigma: I\subset \R\to M$ such that $\frac{\mathrm{d}\sigma}{\mathrm{d}t}(I)\subset C$ and the induced curve $\gamma: \R\to T^*M$, $\gamma=\F\LL^{noh}(\frac{\mathrm{d}\sigma}{\mathrm{d}t})$
verifies that $\frac{\mathrm{d}\gamma}{\mathrm{d}t}(I)\subset \alpha_M^{-1}\lp\Sigma^{noh}\rp$. Locally, $\sigma$ must satisfy the system of equations  (\ref{Noholo}).

An interesting use of Tulczyjew's triple in order to define Lagrangian submanifolds and generalized Legendre transformations within the nonholonomic framework can be found in \cite{Yosi}.

\subsection{Variational constrained equations}\label{VC}
Now, we study the same problem but now using purely variational techniques.
As above, let consider a regular Lagrangian $\mathbb{L}:TM\Flder\R$, and a set of nonholonomic constraints $\phi^{\alpha}(q^{i},\dot q^{i})$, $1\leq\alpha\leq n$, determining a $2m-n$ dimensional submanifold $C\subset TM$. Now we take the extended Lagrangian $\mathcal{L}=\mathbb{L}+\lambda_{\alpha}\phi^{\alpha}$ which includes the Lagrange multipliers $\lambda_{\alpha}$ as new extra variables. The equations of motion for the constrained variational problem are the Euler-Lagrange equations for $\mathcal{L}$, that is:

\begin{eqnarray}\nonumber
&&\frac{\mathrm{d}}{\mathrm{d}t}\lp\frac{\der\mathbb{L}}{\der\dot q^{i}}\rp-\frac{\der\mathbb{L}}{\der q^{i}}=-\dot\lambda_{\alpha}\frac{\der\phi^{\alpha}}{\der\dot q^{i}}-\lambda_{\alpha}\left[\frac{\mathrm{d}}{\mathrm{d}t}\lp\frac{\der\phi^{\alpha}}{\der\dot q^{i}}\rp-\frac{\der\phi^{\alpha}}{\der q^{i}}\right],\\\label{Vako}\\\nonumber
&&\phi^{\alpha}(q^{i},\dot q^{i})=0,\,\,\,\,1\leq\alpha\leq n.
\end{eqnarray}

From a geometrical point of view, these type of variationally constrained problems are determined by a pair $(C, L)$ where $C$ is a submanifold of $TM$, with inclusion $i_C: C\hookrightarrow TM$, and $L: C\to \R$ a Lagrangian function. Using Theorem \ref{tulchi} we deduce that $\Sigma_L$ is a Lagrangian submanifold of $(T^*TM, \omega_{TM})$ (see \cite{GG}). Now using the Tulczyjew's symplectomorphism $\alpha_M$, we induce a new Lagrangian submanifold
$\alpha_M^{-1}\lp\Sigma_L\rp$ of $(TT^*M, \mathrm{d}_T\omega_M)$, which completely determines the constrained variational dynamics.
Of course, the case of unconstrained Lagrangian mechanics is generated taking the whole space $TM$ instead of $C$ and an a  Lagrangian function over the tangent bundle $L: TM\to \R$.

Next we shall prove that, indeed, this procedure gives the correct equations for the constrained variationalpp dynamics.
Take an arbitrary extension $\mathbb{L}:TM\to\R$ of $L: C\to \R$, that is, ${\mathbb L}\circ i_C=L$. As above, assume also that we have fixed local  constraints such that locally determines $C$ by their vanishing, i.e:  $\phi^{\alpha}(q,\dot q)=0$, $1\leq\alpha\leq n$ where
$n=2\hbox{dim } M-\hbox{dim } C$.
		
Locally
\begin{eqnarray}\label{SigmaLCont}
\Sigma_L&=&\{
(q^i, \dot{q}^i, \mu_{i}, \tilde{\mu}_i)\in T^*TM\; |\; \\\nonumber
&&\mu_i=\frac{\partial \mathbb{L} }{\partial q^i} +\lambda_{\alpha}\frac{\partial \phi^{\alpha} }{\partial q^i},\\\nonumber
&&\tilde{\mu}_i=\frac{\partial \mathbb{L} }{\partial \dot{q}^i} +\lambda_{\alpha}\frac{\partial \phi^{\alpha} }{\partial \dot{q}^i},\\\nonumber
&&\phi^{\alpha}(q,\dot q)=0,\,\,\, 1\leq\alpha\leq n\}\; .
\end{eqnarray}
Observe that locally the conormal bundle $\nu^{*}(C)=\hbox{span }\{\mathrm{d}\phi^{\alpha}, 1\leq\alpha\leq n\}$.

Therefore,
\begin{eqnarray}\label{TulcSigmaLCont}
\alpha_M^{-1}\lp\Sigma_L\rp&=&\{(q^i, p_i, \dot{q}^i, \dot{p}_i)\in TT^*M\;|\;\\\nonumber
&&p_i=\frac{\partial \mathbb{L} }{\partial \dot{q}^i} +\lambda_{\alpha}\frac{\partial \phi^{\alpha} }{\partial \dot{q}^i},\\\nonumber
&&\dot{p}_i=\frac{\partial \mathbb{L} }{\partial q^i} +\lambda_{\alpha}\frac{\partial \phi^{\alpha} }{\partial q^i},\\\nonumber
&&\phi^{\alpha}(q,\dot q)=0,\,\,\, 1\leq\alpha\leq n\}\; .
\end{eqnarray}
The solutions for the dynamics given by $\alpha_M^{-1}\lp\Sigma_L\rp\subset TT^*M$ are curves $\gamma: I\subset \R\to T^*M$ such that $\frac{\mathrm{d}\gamma}{\mathrm{d}t}:I\subset \R\to TT^*M$ verifies that $\frac{\mathrm{d}\gamma}{\mathrm{d}t}(I)\subset \alpha_M^{-1}\lp\Sigma_L\rp$. Locally, if $\gamma(t)=(q^i(t), p_i(t))$ then it must verify the following set of differential equations:
\begin{eqnarray*}
\frac{\mathrm{d}}{\mathrm{d}t}\left(\frac{\partial \mathbb{L} }{\partial \dot{q}^i} +\lambda_{\alpha}\frac{\partial \phi^{\alpha} }{\partial \dot{q}^i}\right)-\frac{\partial \mathbb{L} }{\partial q^i} -\lambda_{\alpha}\frac{\partial \phi^{\alpha} }{\partial q^i}&=&0,\\
\phi^{\alpha}(q^{i},\dot q^{i})&=&0,
\end{eqnarray*}
which clearly coincide with equations (\ref{Vako}).

Now, we consider adapted coordinates $(q^i, \dot{q}^a)$ to the submanifold $C$ (recall that $\tau_M(C)=M$ is now a fibration $C\Flder M$), $1\leq i\leq \dim M$ and $1\leq a\leq \dim M-n$, such that
\[
i_C(q^i, \dot{q}^a)=\lp q^i, \dot{q}^a, \Psi^{\alpha}(q^i, \dot{q}^a)\rp.
\]
This means that $\phi^{\alpha}(q^i,\dot q^i)=\dot q^{\alpha}-\Psi^{\alpha}(q^i,\dot q^a)=0$. Therefore, we have
\begin{eqnarray}\label{SigmaLLin}
\Sigma_L&=&\{
(q^i, \dot{q}^i, \mu_{i},  \tilde{\mu}_{i}) |\; \\\nonumber
&&\mu_i=\frac{\partial {L} }{\partial q^i}-\tilde{\mu}_{\alpha}\frac{\partial \Psi^{\alpha} }{\partial q^i}, \\\nonumber
&&\tilde{\mu}_a=\frac{\partial {L} }{\partial \dot{q}^a} -\tilde{\mu}_{\alpha}\frac{\partial \Psi^{\alpha} }{\partial \dot{q}^a},\\\nonumber
&&\dot{q}^{\alpha}=\Psi^{\alpha}(q^i, \dot{q}^a),\,\,\, 1\leq\alpha\leq n\}\; .
\end{eqnarray}
Observe that $(q^i, \dot{q}^a, \tilde{\mu}_{\alpha})$ determines a local system of coordinates for $\Sigma_L$.

Then,
\begin{eqnarray}\label{TulcSigmaLLin}
\alpha_M^{-1}\lp\Sigma_L\rp&=&\{(q^i, p_i, \dot{q}^i, \dot{p}_i)\in TT^*M\;|\\\nonumber
&&p_a=\frac{\partial L }{\partial \dot{q}^a} -p_{\alpha}\frac{\partial \Psi^{\alpha} }{\partial \dot{q}^a},\\\nonumber
&&\dot{p}_i=\frac{\partial {L} }{\partial q^i} -p_{\alpha}\frac{\partial \Psi^{\alpha} } {\partial q^i},\\\nonumber
&&\dot{q}^{\alpha}=\Psi^{\alpha}(q^i, \dot{q}^a),\,\,\, 1\leq\alpha\leq n\}\; .
\end{eqnarray}
Consequently, the solutions must verify the following system of differential equations (see \cite{CLMM}):
 \begin{eqnarray*}
\frac{\mathrm{d}}{\mathrm{d}t}\left(\frac{\partial L }{\partial \dot{q}^a} -p_{\alpha}\frac{\partial \Psi^{\alpha} }{\partial \dot{q}^a}\right)&=&\frac{\partial {L} }{\partial q^a} -p_{\alpha}\frac{\partial \Psi^{\alpha} } {\partial q^a}\\
\dot{p}_{\beta}&=&\frac{\partial {L} }{\partial q^\beta} -p_{\alpha}\frac{\partial \Psi^{\alpha} } {\partial q^\beta},  \\
&&\dot{q}^{\alpha}=\Psi^{\alpha}(q^i, \dot{q}^a),\,\,\, 1\leq\alpha\leq n \; .
\end{eqnarray*}

\subsubsection{The constrained Legendre transformation}

\begin{definition}\label{DefLegTrans}
We define the \emph{constrained Legendre transformation} $\F L: \Sigma_L\longrightarrow T^*M$ as the mapping
 $\F L=\tau_{T^*M}\circ (\alpha^{-1}_M)|_{\Sigma_L}$.

We will say that the constrained system $(L, C)$ is \emph{regular} if $\F L$ is a local diffeomorphism and \emph{hyperregular} if $\F L$ is a global diffeomorphism.
\end{definition}

Observe that locally, if as above we consider the constraints $\dot{q}^{\alpha}=\Psi^{\alpha}(q^i, \dot{q}^a)$ determining  $C$, then
\[
\F L (q^i, \dot{q}^a, \tilde{\mu}_{\alpha})=(q^i, p_a=\frac{\partial L }{\partial\dot{q}^a}-\tilde{\mu}_{\alpha}
 \frac{\partial \Psi^{\alpha} }{\partial \dot{q}^a}, p_{\alpha}=\tilde{\mu}_{\alpha})\; .
\]
The constrained system $(L, C)$ is regular if and only if
$\left(\frac{\partial^2 L}{\partial \dot{q}^a\partial \dot{q}^b}-\tilde{\mu}_{\alpha}\frac{\partial^2 \Psi^{\alpha} }{\partial \dot{q}^a\partial \dot{q}^b} \right)$ is a nondegenerate matrix.

Next, define the \emph{energy function} $E_{L}: \Sigma_L\to \R$ by
\[
E_L(\alpha_u)=\langle \alpha_u, u^V_{u}\rangle-L(u), \quad \alpha_u\in \Sigma_L, u\in C\equiv i_C(C)
\]
Locally, we have
\[E_L(q^i, \dot{q}^a, \tilde{\mu}_{\alpha})=\dot{q}^a\frac{\partial L}{\partial \dot{q}^a}-\tilde{\mu}_{\alpha}\frac{\der\Psi^{\alpha}(q^i, \dot{q}^a)}{\der\dot q^{a}}\,\dot q^{a}+\tilde{\mu}_{\alpha}\Psi^{\alpha}(q^i, \dot{q}^a)-L(q^i, \dot{q}^a).
\]

\begin{remark}
{\rm
The constrained Legendre transformation allows us to develop a Lagrangian formalism on $\Sigma_L$. Indeed,  we can define
the 2-form $\omega_L=(\F L)^*\omega_M$ on $\Sigma_L$ and it is easy to show that the equations of motion of the constrained system are now intrinsically rewritten as
\[
i_X\omega_L=\mathrm{d}E_L.
\]
In consequence, we could develop an intrinsic formalism on the Lagrangian side, that is a  Klein formalism (\cite{Godbillon}, \cite{Grifone}, \cite{Klein}, \cite{dlr}) for constrained systems without using (at least initially) Lagrangian multipliers.
Moreover, notice that the constrained system is regular if and only if $\omega_L$ is a symplectic 2-form on $\Sigma_L$
}
\end{remark}

Then, if the constrained system $(L, C)$ is hyperregular we can define the Hamiltonian function $H: T^*M\to \R$ by
\[
H=E_L\circ\lp\F L\rp^{-1}\;,
\]
and the corresponding Hamiltonian vector field $X_H$ by $i_{X_H}\omega_M=\mathrm{d}H$. In  this particular case we have that
\[
\hbox{Im} X_H=X_H(T^*M)=\beta_M^{-1}(\mathrm{d}H(T^*M))=\alpha_M^{-1}\lp\Sigma_L\rp.
\]
The second equivalence, $X_H(T^*M)=\alpha_M^{-1}\lp\Sigma_L\rp$, will be studied in $\S$ \ref{LHM}. The next diagram summarizes the above discussion:

\[
\xymatrix{
TT^{*}M\ar[rr]^{\alpha_{M}}\ar[dd]_{\tau_{T^{*}M}} & &T^{*}TM\\
        &\alpha^{-1}_{M}(\Sigma_{L})\ar@{^{(}->}[ul] &\\
T^{*}M\ar[dr]_{H} & &\Sigma_{L}\ar[dl]^{E_{L}}\ar[ul]^{\alpha_{M}^{-1}}\ar[ll]^{\F L}\ar@{^{(}->}[uu]\\
&\R&
}
\]

In the singular case, it is necessary to apply the integrability algorithm  to find, if it exists, a subset  where there are consistent solutions of the dynamics (see \cite{GNH}, \cite{GN1}, \cite{GN2}).

\subsection{Comparison of nonholonomic and variational constrained equations. Continuous picture}
Let consider a system defined by the Lagrangian function $\LL:TM\Flder\R$ and an independent set of constraints $\phi^{\alpha}(q^i,\dot q^i)=0$ determining the submanifold $C\subset TM$.

As shown in $\S$ \ref{NC} and $\S$ \ref{VC}, the solutions of the nonholonomic dynamics are geometrically described by the submanifold $\Sigma^{noh}\hookrightarrow TT^*M$, while the solutions of the constrained variational dynamics are given by the Lagrangian submanifold $\Sigma_L\hookrightarrow TT^*M$, where $L=\LL\big|_C:C\Flder\R$.

Our aim is to know whether, given a solution of the nonholonomic problem, it is also a solution of the constrained variational problem. In order to capture the common solutions to both problems, we have developed the following geometric algorithm. Consider the fibered product $T^*M\oplus T^*M$, where we choose the local coordinates $(q^i,p_i,\pi_i)$; consider also the tangent bundle $T(T^*M\oplus T^*M)$ which can be identified with $TT^*M\oplus_{_{T\pi_M}}TT^*M$, which fibers over $TM$. Under these considerations, construct the submanifold $\Sigma^{cons}\hookrightarrow TT^*M\oplus_{_{T\pi_M}}TT^*M$ as follows:
\begin{small}
\begin{eqnarray}
\Sigma^{cons}=&\{&(X_{\alpha_q},Y_{\beta_q})\in TT^*M\oplus_{_{T\pi_M}}TT^*M\, /\, T_{\alpha_q}\pi_M(X_{\alpha_q})=T_{\beta_q}\pi_M(Y_{\beta_q}),\nonumber\\ \,\,&\mbox{for}&\,\, X_{\alpha_q}\in\Sigma^{noh},Y_{\beta_q}\in\Sigma_{L}\}.\label{sij2}
\end{eqnarray}
\end{small}
It is quite clear that the submanifold $\Sigma^{cons}$ gathers together both nonholonomic and constrained variational dynamics. Locally, $\Sigma^{cons}$ is determined by the coordinates $(q^i,p_i,\pi_i,\dot q^i,\dot p_i,\dot\pi_i)$ obeying the nonholonomic and constrained variational conditions respectively presented in \eqref{sij} and \eqref{SigmaLCont}, that is
\[
\begin{array}{lll}
p_i=\frac{\der\LL}{\der\dot q^i}, && \pi_i=\frac{\der\LL}{\der\dot q^i}+\mu_{\alpha}\frac{\der\phi^{\alpha}}{\der\dot q^i},\\\\
\dot p_i=\frac{\der\LL}{\der q^i}+\lambda_{\alpha}\frac{\der\phi^{\alpha}}{\der\dot q^i},&&\dot\pi_i=\frac{\der\LL}{\der q^i}+\mu_{\alpha}\frac{\der\phi^{\alpha}}{\der q^i},
\end{array}
\]
subject to $\phi^{\alpha}(q^i,\dot q^i)=0$. Here, $\lambda_{\alpha}$ and $\mu_{\alpha}$ are the nonholonomic and variational constrained Lagrange multipliers, respectively. Finally, in order to find the common solutions we will need to apply the integrability algorithm described in \cite{CLMM,Mendella}.

The following diagram shows the bundle relations:
\[
\xymatrix{
\Sigma^{cons}\ar@{^{(}->}[r] & TT^*M\oplus_{_{T\pi_M}}TT^*M\ar[rrr]^{_{\widetilde{T\pi_M}}}\ar[dd]_{_{(\tau_{T^*M},\tau_{T^*M})}} & & & TM\ar[dd]^{\tau_M} \\ \\
&T^*M\oplus T^*M\ar[rrr]_{\widetilde{\pi_M}} & && M
}
\]
where $\widetilde{T\pi_M}:TT^*M\oplus_{_{T\pi_M}}TT^*M\longrightarrow TM$ and $\widetilde{\pi_M}:T^*M\oplus T^*M\longrightarrow M$ denote the fibrations of the Whitney sums over $TM$ and $M$, respectively.

As a simple example, consider the case of linear constraints, namely $\phi^{\alpha}(q^i,\dot q^i)$ $=\dot q^{\alpha}-\varphi^{\alpha}_a(q^i)\,\dot q^a=0$. The relationship determining $\Sigma^{cons}$ presented in \eqref{sij2}, as well as the integrability algorithm, implies the following equation:
\begin{equation}\label{curveer}
\dot q^a\mu_{\alpha}R_{ab}^{\alpha}=0,
\end{equation}
where
\[
R^{\alpha}_{ab}=\frac{\der\varphi_b^{\alpha}}{\der q^a}-\frac{\der\varphi_a^{\alpha}}{\der q^b}+\varphi_ a^{\beta}\frac{\der\varphi_b^{\alpha}}{\der q^{\beta}}-\varphi_ b^{\beta}\frac{\der\varphi_a^{\alpha}}{\der q^{\beta}}
\]
can be considered as the curvature of the connection $\Gamma$ in the local projection $\rho(q^a,q^{\alpha})=(q^{\alpha})$ such that the horizontal distribution $\mathcal{H}$ is given by prescribing its annihilator to be
\[
\mathcal{H}^{0}=\{ \mathrm{d}q^{\alpha}-\varphi_a^{\alpha}\mathrm{d}q^{a}\,,\, 1\leq\alpha\leq m\}.
\]
 See more details in \cite{CLMM}.

\subsection{ Lagrangian and Hamiltonian mechanics relationship}\label{LHM}
In this section, we shall discuss the converse case, i.e., starting from a Hamiltonian system we shall show that it is possible to construct a constrained Lagrangian system providing the same dynamics.

Since $\pi_M: T^*M\to M$ is a vector bundle, then it is possible to define the
dilation vector field or Lioville vector field $\Delta^*\in\mathfrak{X}(T^{*}M)$, which is the generator of the one-parameter group of dilations along
the vertical fibres $\mu_q\longrightarrow e^t\mu_q$, $\mu_q\in T^*_q Q$.
The dilation vector field is locally expressed by
\[
\Delta^*=p_i\frac{\partial}{\partial p_i}.
\]

Given a  Hamiltonian  function $H: T^*M\to \R$, the Hamilton's equations are written in canonical coordinates by
\begin{equation}\label{HamVecFi}
\begin{array}{rcl}
\dot{q}^i&=&\displaystyle{\frac{\partial H}{\partial p_i}},\\ \\
\dot{p}_i&=& -\displaystyle{\frac{\partial H}{\partial q^i}}.
\end{array}
\end{equation}
The solutions of the Hamilton's equations are just the integral curves of the Hamiltonian vector field given by
\begin{equation}\label{asd}
i_{X_{H}}\omega_M=\mathrm{d}H,
\end{equation}
where $\omega_{M}$ is a symplectic form on $T^{*}M$ .

Given the Hamiltonian function,  one  defines a function $\F H: T^*M\to TM$, i.e. the fiber derivative of $H$ (see \cite{AbMars}), by
\[
\bra\F H(\alpha_q)\,,\,\beta_q\ket=\frac{\mathrm{d}}{\mathrm{d}t}\Big|_{t=0}H(\alpha_q+t\beta_q),
\]
where both $\alpha_{q},\beta_{q}\in T_q^{*}M$. In local coordinates,
\[
\F H (q^i, p_i)=(q^i, \frac{\partial H}{\partial p_i})\; .
\]

Assume that the image of $T^*M$ under $\F H$ defines a submanifold $C$ of $TM$. Mimicking the Gotay and Nester's definition (\cite{GN1},\cite{GN2}), we implicitly define  the function $L: C\to \R$ by
\begin{equation}\label{lagrangian}
L\circ \F H=\Delta^*H-H.
\end{equation}
The function $L: C\to \R$ will be well-defined if and only if, for any two points $\alpha_{q}, \beta_{q}\in T^*M$ such that $\F H(\alpha_{q})=\F H (\beta_{q})$, we have that $(\Delta^*H-H)(\alpha_{q})=(\Delta^*H-H)(\beta_{q})$. Obviously, without additional assumptions there is no reason why this should be true. The following definition states under what conditions such projection $L$ exists.

\begin{definition}
A Hamiltonian $H: T^*M\to \R$ is {\bf almost-regular} if $\F H (T^*M)=C$ is a submanifold of $TM$ and $\F H: T^*M\to C\subset  TM$ is  a submersion with connected fibers.
\end{definition}

\[
\xymatrix{
T^{*}M\ar[rr]^{\F\,H}\ar[drr] & &TM\\
&&\F H(T^{*}M)=C\ar@{^{(}->}[u]}
\]
Under the assumption that the Hamiltonian $H$ is almost-regular,  it is only necessary to show that expression (\ref{lagrangian}) defines a single-valued function $L: C\to \R$, or, in other words, that $\Delta^{*}H-H$ is a constant function in the fibers. Since each fiber of the submersion $\F H$ is connected, it is sufficient to consider the infinitesimal condition, i.e. to show that
\[
\mathcal{L}_Z (\Delta^*H-H)=0,\,\,\, \hbox{  for all  }\,\,\, Z\in \ker(\F H_*)\;,
\]
where $\mathcal{L}_{Z}$ is the Lie derivate in the $Z$ direction. Working in local coordinates, $Z$ will be of the form
\[
Z=Z_i\frac{\partial }{\partial p_i}, \hbox{  with  }  Z_i\,\frac{\partial^{2} H}{\partial p_i\,\partial p_j}=0, \hbox{  for all  } 1\leq j\leq n\;.
\]
Since $\F H_{*}(Z)=0$, the last condition can be obtained taking into account that $\bra\sigma\,,\,\F H_{*}(Z)\ket$ $=\bra\F H^{*}(\sigma)\,,\,Z\ket=0$, for $\sigma$ an arbitrary point of $T^{*}TM$ such that $\pi_{TM}(\sigma)=\tau_{TM}\lp\F H_{*}(Z)\rp$. Then
\begin{eqnarray*}
&&\mathcal{L}_Z (\Delta^*H-H)= Z_i\frac{\partial\lp\Delta^*H-H\rp}{\partial p_i}\\
&&=Z_{i}\frac{\partial H}{\partial p_i}+p_j Z_{i}\,\frac{\partial^{2} H}{\partial p_i\,\partial p_j}-Z_{i}\,\frac{\partial H}{\partial p_i}=0.
\end{eqnarray*}
In what follows we will assume that $H$ satisfies the almost regularity property.

 \begin{theorem}\label{TulczyAlpha}
 The following equality holds
 \[
 \alpha_M(X_H(T^*M))=\Sigma_{L}\; ,
 \]
where $\alpha_{M}$ is the Tulczyjew's isomorphism.
  \end{theorem}
\begin{proof}
Take $W_1\in \Sigma_{X_H}=X_{H}(T^{*}M)\subset TT^*M$ and take $\alpha_M(W_1)\in T^*TM$.  We need to prove that
\[
\left\langle\alpha_M(W_1), U\right\rangle = \left\langle
        \mathrm{d} L , U \right\rangle,
\]
for all  $U \in T\,C \subset TTM$, such that $\tau_{TM} (U) = T\pi_M(W_1)$. This is equivalent to the equality
\begin{equation}\label{Pr1}
\left\langle\alpha_M(W_1)\,,\,\F H_*(W_2) \right\rangle = \left\langle
\mathrm{d} L\,,\,\F H_*(W_2) \right\rangle,
\end{equation}
for all $W_2\in TT^*M$ such that $\tau_{TM}\lp\F H_*(W_{2})\rp = T\pi_M(W_1)$. Therefore, regarding (\ref{Pr1}) the previous equality  turns out to be
\begin{equation}\label{Pr2}
(\F H)^*\lp\alpha_M(W_1)\rp=(\F H)^*\lp \mathrm{d}L\rp=\mathrm{d}(\Delta^* H-H)\;,
\end{equation}
where the right hand of the equation comes directly from (\ref{lagrangian}).
If locally $W_1=\dot{q}^i\frac{\partial}{\partial q^i}+ \dot{p}_i\frac{\partial}{\partial p_i}$, in other words
$W_1=(q^i, p_i; \dot{q}^i, \dot{p}_i)$, then
\begin{eqnarray}\nonumber
(\F H)^*\lp\alpha_M(W_1)\rp&=&\left(\dot{p}_i + p_j\frac{\partial^{2} H}{\partial p_j\partial q^i}\right)\, dq^i+p_j\frac{\partial^{2} H}{\partial p_j\partial p_i}\, d{p}_i\\\label{Poof}
&=&\left(p_j\frac{\partial^{2} H}{\partial p_j\partial q^i}-\frac{\partial H}{\partial q^i}\right)\, dq^i
+p_j\frac{\partial^{2} H}{\partial p_j\partial p_i}\, d{p}_i,
\end{eqnarray}
where we consider $\dot{p}_i=-\frac{\partial H}{\partial q^i}$ since we are dealing with a Hamiltonian vector field $X_{H}$ (see equations (\ref{HamVecFi})). From condition,  $\tau_{TM}\F H_*(W_{2}) = T\pi_M(W_1)$ we also deduce that
$\dot{q}^i=\frac{\partial H}{\partial p_i}$. But this is true since $W_1\in \Sigma_{X_H}$.

On the other hand, a straightforward computation leads to check that $\mathrm{d}(\Delta^{*}H-H)$ is exactly (\ref{Poof}).
\end{proof}

%
%


\section{Discrete equivalence}\label{DEQ}

\subsection{Discrete nonholonomic mechanics}\label{ND}

A discrete nonholonomic system is determined by three ingredients: a discrete lagrangian $\LL_d: M\times M\rightarrow \R$, a constraint distribution ${\mathcal D}$ on $M$ and a discrete constraint submanifold $C_d\subset M\times M$ with canonical inclusion $i_{C_d}: C_d\hookrightarrow M\times M$. Notice that discrete mechanics can also be seen within this case, where ${\mathcal D}=TM$ and $C_d=M\times M$. Notice also that, in the discrete version of Lagrangian mechanics, the tangent manifold $TM$ is substituted by the cartesian product $M \times M$.

Define the  affine submanifold  $\Sigma_{d}^{noh}\subset T^*(M\times M)$ by
$\Sigma_{d}^{noh}=(\mathrm{d}\LL_d)\circ i_{C_d}+F_d$,
 where $F_d$ is the vector subbundle of $T_{C_d}^*(M\times M)$ given by
 \[
 F_d=\left(pr_1^*\,{\mathcal D}^0\right)\big|_{C_d}\; .
 \]
Here, $pr_1: M\times M\rightarrow M$ is the first projection onto $M$.

The symplectic manifold $(T^*(M\times M)\,,\,\omega_{M\times M})$ is
symplectomorphic to $(T^*M\times T^*M\,,\,\Omega)$, where
$\Omega={pr}_1^*\omega_M-{pr}_0^{*}\omega_M$ and ${pr}_i:T^*M\times T^*M\rightarrow T^{*}M$, $i=0,1$, are the natural projections of $T^*M\times T^*M$ onto $T^*M$. The symplectomorphism is given by
\[
\begin{array}{rcl}
\Upsilon:T^* (M\times M)&\to &T^* M\times T^* M\\
\gamma_{(q_0, q_1)}\equiv (\gamma _{q_0},\gamma _{q_1})&\mapsto &(-\gamma _{q_0},\gamma
_{q_1})
\end{array}
\]
where $(q_0, q_1)\in M\times M$.

Using $\Upsilon$ we induce the affine  subbundle $\Upsilon(\Sigma_{d}^{noh})$ of the symplectic manifold $(T^*M\times T^*M, \Omega)$.
The dynamics is then determined by the sequences $\gamma_{q_0}, \gamma_{q_1}, \ldots, \gamma_{q_N}$ such that $(\gamma_{q_i}, \gamma_{q_{i+1}})\in \Upsilon(\Sigma^{noh}_d)$, $0\leq i\leq N-1$ (see \cite{IMMP}).

We will describe now the equations in terms of local coordinates.
Assume that $C_d$ is defined by the vanishing of the following set of independent constraints:  $\phi_d^{\alpha}(q_0, q_1)=0$, $1\leq\alpha\leq n$, where
$n=2\hbox{dim } M-\hbox{dim } C$ and ${\mathcal D^0}=\hbox{span} \{\omega^{\alpha}=\omega^{\alpha}_i\, \mathrm{d}q^i\}$.
Therefore,
\begin{eqnarray}\label{SigmaL-noh}
\Sigma_{d}^{noh}&=&\{
(q^i_0, {q}_1^i, (\mu_{0})_{i}, (\mu_{1})_i)\in T^*(M\times M)\; |\; \\\nonumber
&&(\mu_{0})_i=\frac{\partial \LL_d }{\partial q_0^i} +(\lambda_{0})_{\alpha}\,\omega^{\alpha}_i,\\\nonumber
&&(\mu_{1})_i=\frac{\partial \LL_d }{\partial {q}_1^i}, \\\nonumber
&&\phi_d ^{\alpha}(q_0, q_1)=0,\,\,\, 1\leq\alpha\leq n\}\;,
\end{eqnarray}
where $(\lambda_{\alpha})_0$ are Lagrange multipliers to be determined.
Then, \begin{eqnarray}\label{UpSigmaL-noh}
\Upsilon(\Sigma_{d}^{noh})&=&\{(q_0^i, (p_{0})_i, {q}_1^i, (p_{1})_i)\in T^*M\times T^*M\;|\;\\\nonumber &&(p_{0})_i=-\frac{\partial \mathbb{L}_d }{\partial q_0^i} -(\lambda_{0})_{\alpha}\,\omega^{\alpha}_i,\\\nonumber
&&(p_{1})_i=\frac{\partial \mathbb{L}_d }{\partial q_1^i} ,\\\nonumber
&&\phi_d^{\alpha}(q_0, q_1)=0,\,\,\,\, 1\leq\alpha\leq n\}\; .
\end{eqnarray}
The solutions of the dynamics are therefore  given by the equations
\begin{small}
\begin{eqnarray*}
&&\frac{\partial \mathbb{L}_d }{\partial q_1^i}(q_{k-1}, q_k)=-\frac{\partial \mathbb{L}_d }{\partial q^i_{0}}(q_k, q_{k+1}) -(\lambda_{k})_{\alpha}\,\omega^{\alpha}_i(q_k),\\
&&\phi_d^{\alpha}(q_k, q_{k+1})=0.
\end{eqnarray*}
\end{small}
 These equations are traditionally written in the following manner
\begin{eqnarray*}
D_2\mathbb{L}_d (q_{k-1}, q_k)+D_1\mathbb{L}_d (q_{k}, q_{k+1})&=&
(\lambda_{k})_{\alpha}\,\omega^{\alpha}(q_k)\\
\phi_d^{\alpha}(q_k, q_{k+1})&=&0,
\end{eqnarray*}
which are the expression of the discrete nonholonomic equations  (see \cite{CoSMa} for more details).

\subsection{Constrained discrete Lagrangian mechanics}\label{VD}

A discrete cons\-trai\-ned system \cite{MMS} is determined by a pair  $(C_d, L_d)$ where $C_d$ is a submanifold of $M\times M$, with inclusion $i_{C_d}: C_d\hookrightarrow M\times M$, and $L_{d}: C_d\to \R$ a discrete Lagrangian function.

Using again Theorem \ref{tulchi} we deduce   that $\Sigma_{L_d}$ is a Lagrangian submanifold of $(T^*(M\times M)\,,\,\omega_{M\times M})$.

Using $\Upsilon$ we induce the Lagrangian submanifold $\Upsilon(\Sigma_{L_d})$ of the symplectic manifold $(T^*M\times T^*M, \Omega)$ (see the diagram below).
\[
\xymatrix{
T^{*}(M\times M)\ar[rr]^{\Upsilon} & &T^{*}M\times T^{*}M\\
\Sigma_{L_{d}}\ar[rr]^{\Upsilon}\ar@{^{(}->}[u] & &\Upsilon(\Sigma_{L_{d}})\ar@{^{(}->}[u]
}
\]
The dynamics is determined by the sequences $\gamma_{q_0}, \gamma_{q_1}, \ldots, \gamma_{q_N}$ such that $(\gamma_{q_i}, \gamma_{q_{i+1}})\in \Upsilon(\Sigma_{L_d})$, $0\leq i\leq N-1$ (see \cite{IMMP,MMS}).
Observe that
\begin{equation}\label{CondicionDinamica}
\gamma_{q_k}\in T_{q_k}^*M\cap pr_0(\Upsilon(\Sigma_{L_d}))\cap pr_1(\Upsilon(\Sigma_{L_d})), 1\leq k\leq N-1.
\end{equation}

After determining intrinsically the dynamics, as in the continuous case, we now consider local expressions.
Take an arbitrary extension $\mathbb{L}_d: M\times M\to\R$ of $L_d: C_d\to \R$, that is, ${\mathbb L}_d\circ i_{C_d}=L_d$. Assume also that we have fixed local  constraints such that determines the submanifold $C_d$. This definition is performed by the vanishing of the following set of independent constraints:  $\phi_d^{\alpha}(q_0, q_1)=0$, $1\leq\alpha\leq n$ where
$n=2\hbox{dim } M-\hbox{dim } C$.
		
Locally
\begin{eqnarray}\label{SigmaL}
\Sigma_{L_d}&=&\{
(q^i_0, {q}_1^i, (\mu_{0})_{i}, (\mu_{1})_i)\in T^*(M\times M)\; |\; \\\nonumber
&&(\mu_{0})_i=\frac{\partial \mathbb{L}_d }{\partial q_0^i} +(\lambda_{1})_{\alpha}\frac{\partial \phi_d^{\alpha} }{\partial q_0^i},\\\nonumber
&&(\mu_{1})_i=\frac{\partial \mathbb{L}_d }{\partial {q}_1^i} +(\lambda_{1})_{\alpha}\frac{\partial \phi_d^{\alpha} }{\partial {q}_1^i},\\\nonumber
&&\phi_d ^{\alpha}(q_0, q_1)=0,\,\,\, 1\leq\alpha\leq n\}\;,
\end{eqnarray}
where $(\lambda_{1})_{\alpha}$ are Lagrange multipliers to be determined.

Therefore,
\begin{eqnarray}\label{UpSigmaL}
\Upsilon(\Sigma_{L_d})&=&\{(q_0^i, (p_{0})_i, {q}_1^i, (p_{1})_i)\in T^*M\times T^*M\;|\;\\\nonumber &&(p_{0})_i=-\frac{\partial \mathbb{L}_d }{\partial q_0^i} -(\lambda_{1})_{\alpha}\frac{\partial \phi^{\alpha}_d }{\partial q_0^i},\\\nonumber
&&(p_{1})_i=\frac{\partial \mathbb{L}_d }{\partial q_1^i} +(\lambda_{1})_{\alpha}\frac{\partial \phi_d^{\alpha} }{\partial q_1^i},\\\nonumber
&&\phi_d^{\alpha}(q_0, q_1)=0,\,\,\,\, 1\leq\alpha\leq n\}\; .
\end{eqnarray}
The solutions of the dynamics come from (\ref{CondicionDinamica}) and are given by the equations
\begin{small}
\begin{eqnarray*}
&&\frac{\partial \mathbb{L}_d }{\partial q_1^i}(q_{k-1}, q_k) +(\lambda_{k})_{\alpha}\frac{\partial \phi_d^{\alpha} }{\partial q_1^i}(q_{k-1}, q_{k})=-\frac{\partial \mathbb{L}_d }{\partial q^i_{0}}(q_k, q_{k+1}) -(\lambda_{k+1})_{\alpha}\frac{\partial \phi^{\alpha}_d }{\partial {q}_0^i}(q_k, q_{k+1}),\\
&&\phi_d^{\alpha}(q_{k-1}, q_{k})=0,\\
&&\phi_d^{\alpha}(q_k, q_{k+1})=0.
\end{eqnarray*}
\end{small}
 These equations are traditionally written as
\begin{eqnarray*}
&&D_2\mathbb{L}_d (q_{k-1}, q_k)+D_1\mathbb{L}_d (q_{k}, q_{k+1})\\
&&+ (\lambda_{k})_{\alpha}D_2 \phi_d^{\alpha} (q_{k-1}, q_{k+1})+(\lambda_{k+1})_{\alpha}D_1 \phi^{\alpha}_d(q_{k}, q_{k+1})=0,\\
&&\phi_d^{\alpha}(q_{k-1}, q_{k})=0,\  \quad \phi_d^{\alpha}(q_k, q_{k+1})=0,
\end{eqnarray*}
which are the expression of the discrete vakonomic equations (see \cite{Benito} for more details).

Now, as a particular case, we assume that we can choose  adapted coordinates $(q_0^i, q_1^a)$, $1\leq i\leq \dim M$ and $1\leq a\leq \dim M-n$, on $C_d$ in such a way the  inclusion is written as
\begin{equation}\label{Coor}
i_{C_d}(q_0^i, {q}_1^a)=(q_0^i, {q}_1^a, \Psi_d^{\alpha}(q_0^i, {q}_1^a)).
\end{equation}
In other words, we can write the constraints as $\phi^{\alpha}_{d}(q_{0},q_{1})=q_1^{\alpha}-\Psi_d^{\alpha}(q_0^i, {q}_1^a)=0$.
Thus, locally we have that
\begin{eqnarray}\label{SigmaLd}
\Sigma_{L_d}&=&\{
(q_0^i, {q}_1^i, (\mu_{0})_{i}, (\mu_{1})_{i})\in T^*(M\times M) |\; \\\nonumber
&&(\mu_{0})_i=\frac{\partial {L}_d }{\partial q_0^i}-(\mu_{1})_{\alpha}\frac{\partial \Psi_d^{\alpha} }{\partial q_0^i}, \\\nonumber
&&(\mu_{1})_a=\frac{\partial {L}_d }{\partial {q}_1^a} -(\mu_{1})_{\alpha}\frac{\partial \Psi_d^{\alpha} }{\partial {q}_1^a}, \\\nonumber
&& q_{1}^{\alpha}=\Psi_d^{\alpha}(q_0^i, {q}_1^a),\,\,\,1\leq \alpha\leq n\}\; .
\end{eqnarray}
Observe that $(q_0^i, {q}_1^a, (\mu_{1})_{\alpha})$ gives a local coordinate system  for $\Sigma_{L_d}$.

Therefore, we obtain the following expression of the Lagrangian submanifold $\Upsilon(\Sigma_{L_d})$:
\begin{eqnarray}\label{UpsSigmaLd}
\Upsilon(\Sigma_{L_d})&=&\{(q_0^i, (p_{0})_i, q_1^i, (p_{1})_i)\in T^*M\times T^*M\;|\;\\\nonumber
&&(p_{0})_i=-\frac{\partial L_d }{\partial {q}_0^i} +(p_{1})_{\alpha}\frac{\partial \Psi_d^{\alpha} }{\partial {q}_0^i},\\\nonumber
&&(p_{1})_a=\frac{\partial {L}_d }{\partial q_1^a}-(p_{1})_{\alpha}\frac{\partial \Psi^{\alpha}_d } {\partial q_1^a},\\\nonumber
&&q_{1}^{\alpha}=\Psi_d^{\alpha}(q_0^i, {q}_1^a),\,\,\,  1\leq \alpha\leq n \}\; .
\end{eqnarray}
Consequently, the solutions must verify the following system of difference equations:
 \begin{eqnarray*}
&&(D_2)_{a}( {L}_d-(p_{k})_{\alpha}\Psi^{\alpha}_d)(q_{k-1}, q_k)+(D_1)_{a} ({L}_d-(p_{k+1})_{\alpha}\Psi^{\alpha}_d)(q_{k}, q_{k+1})=0,\\
&&(p_{k})_{\beta}+(D_1)_{\beta}{L}_d(q_{k}, q_{k+1})-(p_{k+1})_{\alpha}(D_1)_{\beta}\Psi^{\alpha}_d(q_{k}, q_{k+1})=0,\\
&&{q}_{k+1}^{\alpha}=\Psi_d^{\alpha}(q_k^i, {q}_{k+1}^a)\; ,
\end{eqnarray*}
where $(D_{j})_{l}$ just means $\frac{\der}{\der q_{j}^{l}}$, being $j=\lc 1,2\rc$ and $l=\lc a,\alpha\rc$.
\subsubsection{The constrained discrete Legendre transformations}

\begin{definition}\cite{MMS}
We define the \emph{constrained discrete Legendre transformations} $\F L_d^\pm: \Sigma_{L_d}\longrightarrow T^*M$ as the mappings
\begin{eqnarray*}
 \F L_d^{-}&=&pr_0\circ (\Upsilon)\big|_{\Sigma_{L_d}},\\
 \F L_d^{+}&=&pr_1\circ (\Upsilon)\big|_{\Sigma_{L_d}}.
 \end{eqnarray*}
We will say that the constrained system $(L_d, C_d)$ is \emph{regular} if $\F L_d^{-}$ is a local diffeomorphism and \emph{hyperregular} if $\F L_d^{-}$ is a  diffeomorphism.
\end{definition}
\begin{remark}
{\rm
It is easy to prove that $\F L_d^{-}$ is a local diffeomorphism if and only if $\F L_d^{+}$ is a local diffeomorphism; therefore, it is possible to characterize the regularity of the constrained system using any of the two Legendre transformations (see \cite{MMS}).
}
\end{remark}

Observe that if we consider the local constraints ${q}_1^{\alpha}=\Psi_d^{\alpha}(q_0^i, {q}_1^a)$ determining  $C_d$, then
\begin{eqnarray*}
\F L_d^{-} (q_0^i, {q}_1^a, (\mu_1)_{\alpha})&=&(q_0^i, (p_0)_i=-\frac{\partial L_d }{\partial {q}_0^i}+(\mu_1)_{\alpha}\frac{\partial \Psi_d^{\alpha} }{\partial q_0^i})\; \\
&=&(q_0^i, -(D_{1})_{i}(L_d-(\mu_1)_{\alpha} \Psi_d^{\alpha})(q_0, q_1))\;.\\
\F L_d^{+} (q_0^i, q_1^a, (\mu_1)_{\alpha})&=&(q_1^a\,,\,q_{1}^{\alpha}=\Psi_d^{\alpha}(q_0^i, {q}_1^a),\\ &&(p_1)_a=\frac{\partial L_d }{\partial {q}_1^a}-(\mu_1)_{\alpha}\frac{\partial \Psi_d^{\alpha} }{\partial {q}_1^a}\,,\,(p_1)_{\alpha}=(\mu_1)_{\alpha})\; \\
&=&(q_1^a\,,\,q_{1}^{\alpha}=\Psi_d^{\alpha}(q_0^i, {q}_1^a), \\
&&(p_1)_a=(D_2)_a(L_d-{\mu}^1_{\alpha}\Psi_d^{\alpha})(q_0, q_1)\,,\,(p_1)_{\alpha}=(\mu_1)_{\alpha})\; .
\end{eqnarray*}
So, the constrained system $(L_d, C_d)$ is regular if and only if the matrix
$$
(A_{ia}, A_{i\alpha})=\left(
\frac{\partial^2 L_d}{\partial {q}_0^i\partial {q}_1^a}-{\mu}^1_{\alpha}\frac{\partial^2 \Psi_d^{\alpha}}{\partial q^i_0\partial q_1^a},\frac{\partial \Psi_d^{\alpha} }{\partial q_0^i} \right)$$ is nondegenerate.

\begin{remark}
{\rm
The constrained Legendre transformations allow us to define a univocally  presymplectic 2-form on $\Sigma_{L_{d}}$
\[
\omega_{L_d}=(\F L_d^-)^*\omega_M= (\F L_d^+)^*\omega_M
\]
which is symplectic if the system is regular.
}
\end{remark}

Then, if the constrained system $(L_d, C_d)$ is hyperregular, we can define the discrete dynamics determined by $\Upsilon^{-1}(\Sigma_{L_{d}})$ as the graph of the canonical transformation $(\F L_d^+)\circ (\F L_d^-)^{-1}: T^*M\to T^*M$, that is,
\[
\hbox{Graph } (\F L_d^+)\circ (\F L_d^-)^{-1}=\Upsilon(\Sigma_{L_d})\; .
\]

\subsection{Comparison of nonholonomic and variational constrained equations. Discrete picture}
Let consider the system defined by the discrete Lagrangian function $\LL_d:M\times M\Flder\R$ and a set of constraints $\phi_d^{\alpha}(q_0,q_1)=0$ determining the submanifold $C_d\subset M\times M$.

As shown in $\S$ \ref{ND} and $\S$ $\ref{VD}$, the solutions of the discrete nonholonomic dynamics are geometrically described by the affine subbundle $\Upsilon(\Sigma_d^{noh})\subset T^*M\times T^*M$, while the solutions of the discrete constrained variational dynamics are given by the Lagrangian submanifold $\Upsilon(\Sigma_{L_d})\subset T^*M\times T^*M$, where $L_d=\LL_d\big|_{C_d}:C_d\Flder\R.$

Given a solution of the discrete nonholonomic problem, we want to know when is also a solution of the associated discrete constrained variational problem.
To capture the set of common solutions to both problems, we have developed
the following geometric integrability algorithm similar to the continuous one.

As in the continuous case, take the  Whitney sum  $T^*M\oplus T^*M$, and the cartesian product
\[(T^*M\oplus T^*M)\times (T^*M\oplus T^*M)\equiv
 (T^*M\times T^*M)\oplus_{\pi_M\times \pi_M} (T^*M\times T^*M).
 \]
Construct  the submanifold $\Sigma^{cons}_d\hookrightarrow (T^*M\oplus T^*M)\times (T^*M\oplus T^*M)$ as follows:
\begin{small}
\begin{eqnarray}
\Sigma^{cons}_d=&\{&(\gamma_{q_0}, \gamma_{q_1}, \widetilde{\gamma}_{q_0}, \widetilde{\gamma}_{q_1})\in (T^*M\times T^*M)\oplus_{\pi_M\times \pi_M} (T^*M\times T^*M) \, /\, \nonumber\\ \,\,&&\,\, (\gamma_{q_0}, \gamma_{q_1})\in\Sigma_d^{noh}, (\widetilde{\gamma}_{q_0}, \widetilde{\gamma}_{q_1})\in\Sigma_{L_d}\}.\label{sij2}
\end{eqnarray}
\end{small}
It is quite clear that the submanifold $\Sigma^{cons}_d$ gathers together both nonholonomic and constrained variational dynamics and applying the discrete version of the integrability algorithm developed in \cite{IMMP} we will find the set where there are common solutions for both dynamics.

The following diagram shows the bundle relations.

\[
\xymatrix{
\Sigma_d^{cons}\ar@{^{(}->}[r] & (T^*M\oplus T^*M)\times (T^*M\oplus T^*M)\ar[rrr]_{\qquad \widetilde{\pi_M}\times \widetilde{\pi_M}}\ar@<1ex>[dd]\ar@<-1ex>[dd] & & & M\times M\ar@<1ex>[dd]\ar@<-1ex>[dd] \\ \\
&T^*M\oplus T^*M\ar[rrr]_{\widetilde{\pi_M}} & && M
}
\]
where the vertical arrows represent the projections onto the first and second factor of the respective cartesian products.

%
\subsection{Construction of a discrete constrained problem from a Lagran\-gian submanifold of $T^*M\times T^*M$}

Given a Lagrangian submanifold $\Lambda$ of $(T^*M\times T^*M, \Omega)$ we will construct, under some regularity conditions, a constrained Lagrangian problem given by a submanifold $C_d\subset M\times M$ and a function $L_d: C_d\to \R$.

We first construct the Lagrangian submanifold $\Upsilon^{-1}(\Lambda)$ of
$(T^*(M\times M), \linebreak \omega_{M\times M})$ and we assume that the restriction of $\omega_{M\times M}$ to $\Upsilon^{-1}(\Lambda)$ is exact, that is, we have a generating function $S: \Upsilon^{-1}(\Lambda)\to \R$. In addition, we suppose that the image of $\Upsilon^{-1}(\Lambda)$ by $\pi_{M\times M}$ is a submanifold $C_d$ of $M\times M$, and, also, that $(\pi_{M\times M})\big|_{\Upsilon^{-1}(\Lambda)}$ is a submersion with connected fibers (see the diagrams below).

\[
\xymatrix{
T^{*}(M\times M)\ar[rr]^{\Upsilon} & &T^{*}M\times T^{*}M\\
\Upsilon^{-1}(\Lambda)\ar@{^{(}->}[u] &&\Lambda\ar@{^{(}->}[u]\ar[ll]^{\Upsilon^{-1}}
}
\]
\vspace{0.2cm}

\[
\xymatrix{
&T^*(M\times M)\ar[dl]_{\pi_{M\times M}}&\\
M\times M\supset C_{d}\ar[dr]_{L_{d}}& &\Upsilon^{-1}(\Lambda)\ar@{^{(}->}[ul]^{i_{\Upsilon^{-1}(\Lambda)}}\ar[dl]^{S}\\
&\R&
}
\]

\begin{theorem}\label{Discreto} Under the previous conditions the function $S:\Upsilon^{-1}(\Lambda)\to \R$  is $(\pi_{M\times M})\big|_{\Upsilon^{-1}(\Lambda)}$-projectable onto a function $L_d: C_d\to \R$. Moreover, the following equation holds
\[
\Upsilon^{-1}(\Lambda)=\Sigma_{L_d}\; .
\]
\end{theorem}

\begin{proof}
The submanifold $\Ups$ is defined as
\begin{equation}\label{ups}
\Upsilon^{-1}(\Lambda)=\{(\gamma_{q_0}, \gamma_{q_1})\in T^*_{(q_{0},q_{1})}(M\times M)\; |\; i^{*}_{_{\Upsilon^{-1}(\Lambda)}}\theta_{M\times M} (\gamma_{q_0}, \gamma_{q_1})=\mathrm{d}S(\gamma_{q_0}, \gamma_{q_1})\}.
\end{equation}
By definition of the Liouville 1-form $\theta_{M\times M}$ we have that
\[
\langle \theta_{M\times M}\,,\,\hbox{ker}\,T\,\pi_{M\times M}\rangle=0.
\]
By applying the chain rule $T(\pi_{M\times M}\circ i_{_{\Ups}})=T\,\pi_{M\times M}\circ T\,i_{_{\Ups}}$, is easy to check that $T\,i_{_{\Ups}}\left(
\hbox{ker}\,T\,\pi_{M\times M}\big|_{_{\Ups}}\right)\subset \hbox{ker}\,T\,\pi_{M\times M}$.
Thus, we finally deduce that $S$ is projectable into $L_d: C_d\to\R$, i.e.
\begin{equation}\label{Ss}
S=(\pi_{M\times M}\big|_{_{\Ups}})^{*}L_{d}
\end{equation}
since $\pi_{M\times M}(\Ups)=C_{d}$.
\vspace{0.2cm}

On the other hand, we can define a tangent vector $X=(X_{\gamma_{q_{0}}},X_{\gamma_{q_{1}}})\in T\,\Ups\subset TT^{*}(M\times M)$, such that $T_{_{(\gamma_{q_{0}},\gamma_{q_{1}})}}\pi_{M\times M}(X)=(u_{q_{0}},u_{q_{1}})\in T_{(q_{0},q_{1})}C_{d}$, by
\[
\bra(\gamma_{q_{0}},\gamma_{q_{1}})\,,\,(u_{q_{0}},u_{q_{1}})\ket=\bra\mathrm{d}S(\gamma_{q_{0}},\gamma_{q_{1}})\,,\,X\ket,
\]
where $(\gamma_{q_{0}},\gamma_{q_{1}})\in\Ups$. The equation above comes directly from the definitions of both the Liouville one-form and the Lagrangian submanifold $\Ups$ in (\ref{ups}). Regarding equation (\ref{Ss}) and taking into account that the pullback and the exterior derivative commute, we arrive to
\begin{eqnarray*}
\bra\mathrm{d}S(\gamma_{q_{0}},\gamma_{q_{1}})\,,\,X\ket&=&\bra\mathrm{d}L_{d}(q_{0},q_{1})\,,\,T\pi_{M\times M}\big|_{_{\Ups}}(X)\ket\\&=&\bra\mathrm{d}L_{d}(q_{0},q_{1})\,,\,(u_{q_{0}},u_{q_{1}})\ket.
\end{eqnarray*}
In the last line of the expression just above, we recognize the definition given in Theorem \ref{tulchi} of $\Sigma_{L_{d}}$, that is

\begin{multline*}
\Sigma_{L_{d}}=\bigl\{(\gamma_{q_{0}},\gamma_{q_{1}})\in T_{(q_{0},q_{1})}^{*}(M\times M)\,|\,\,\bra(\gamma_{q_{0}},\gamma_{q_{1}})\,,\,(u_{q_{0}},u_{q_{1}})\ket=\\
=\bra\mathrm{d}L_{d}(q_{0},q_{1})\,,\,(u_{q_{0}},u_{q_{1}})\ket\,\,\,\mbox{for all}\,\,\,(u_{q_{0}},u_{q_{1}})\in T_{(q_{0},q_{1})}C_{d}\bigr\}.
\end{multline*}
In consequence, we deduce that $\Ups=\Sigma_{L_{d}}$.


\end{proof}

\begin{remark}
{\rm
We would like to point out that the proof of Theorem \ref{Discreto} can be easily extended  to the continuous case. Namely, let consider $\tilde\Lambda\subset TT^{*}M$ a Lagrangian submanifold. Under some regularity conditions, it is possible to construct a constrained Lagrangian problem given by a submanifold $C\subset TM$ and a Lagrangian function $L:C\Flder\R$. If we consider the Lagrangian submanifold $\alpha_{M}(\tilde\Lambda)\subset(T^{*}TM\,,\,\omega_{TM})$, where $\alpha_{M}$ is the Tulczyjew's isomorphism, we can build an analogy with the discrete case by assuming that the restriction of $\omega_{TM}$ to $\alpha_{M}(\tilde\Lambda)$ is exact, that is, we have a generating function $\tilde S:\alpha_{M}(\tilde\Lambda)\Flder\R$;
\[
\alpha_{M}(\tilde\Lambda)=\lc\gamma\in T^{*}_{v_{q}}TM\,|\,i^{*}_{\alpha_{M}(\tilde\Lambda)}\theta_{TM}(\gamma)=\mathrm{d}\tilde S(\gamma)\rc,
\]
where $v_{q}\in TM$ such that $\tau_{_{M}}(v_{q})=q\in M$. In addition, we suppose that $\pi_{_{TM}}\lp\alpha_{M}(\tilde\Lambda)\rp$ is a submanifold $C\subset TM$, and that $\pi_{_{TM}}\big|_{\alpha_{M}(\tilde\Lambda)}$ is a submersion with connected fibers.

Again, by the definition of the Liouville one-form $\theta_{TM}$ we have that $\bra\theta_{_{TM}}\,,\,\mbox{ker}T\pi_{_{TM}}\ket=0$, and consequently that $\tilde S$ is projectable into $L$, that is $\tilde S=(\pi_{_{TM}}\big|_{\alpha_{M}(\tilde\Lambda)})^{*}L$.

Following similar arguments that in the proof of Theorem \ref{Discreto}  we deduce that $\alpha_{M}(\tilde\Lambda)=\Sigma_{L}$.

Since we are not fixing the Lagrangian submanifold $\tilde\Lambda$, this is a more general result than that one provided by Theorem \ref{TulczyAlpha}, i.e. $\alpha_{M}(X_{H}(T^{*}M))=\Sigma_{L}$. Nevertheless, among all the Lagrangian submanifolds $\tilde\Lambda$ of $T^{*}TM$, we choose $X_{H}(T^{*}M)$, that is the image of the cotangent bundle $T^{*}M$ by the Hamiltonian vector field provided by the equations $i_{X_{H}}\omega=\mathrm{d}H$, in order to stress the relationship between  Hamiltonian and  constrained Lagrangian systems.
}

\end{remark}

\section{Examples}

\subsection{Linear Constraints}
Let consider a dynamical system denoted by the Lagrangian
\begin{equation}\label{Lag}
\mathbb{L}(v_{q})=\frac{1}{2}\,g(v_{q},v_{q})-V(q)=\frac{1}{2}g_{ij}\dot{q}^i\dot{q}^j-V(q)\,
\end{equation}
where $v_{q}\in T_qM$, with  local coordinates $v_{q}=(q^i,\dot q^i)$, $g$ is a Riemannian metric with components $(g_{ij})$. Moreover, $V: M\rightarrow \R$ is a potential function. Additionally,  the system is subject to the linear constraints
\begin{equation}\label{Cons}
\phi^{\alpha}(v_{q})=\dot q^{\alpha}-\Gamma^{\alpha}_{a}(q)\,\dot q^{a},
\end{equation}
where $\dot q^{i}=\lc\dot q^{a},\dot q^{\alpha}\rc$.
Locally, the constraints define a submanifold $C\subset TM$. Moreover, we have the restriction of $\mathbb{L}$ to $C$, $L: C\rightarrow \R$. In local coordinates,
\[
L(q^i, \dot{q}^a)=\frac{1}{2}\gamma_{ab}\dot{q}^a\dot{q}^b-V(q)\;,
\]
where
\[
\gamma_{ab}\,(q)=g_{ab}+g_{a\alpha}\Gamma_{b}^{\alpha}(q)+g_{b\alpha}\Gamma_{a}^{\alpha}(q)+g_{\alpha\beta}\Gamma^{\alpha}_{a}(q)\Gamma^{\beta}_{b}(q).
\]
Observe that $(\gamma_{ab})$ is invertible since $g$ is a Riemannian metric.

Using expression  (\ref{TulcSigmaLLin}), we can find local coordinates for
$\alpha_{M}^{-1}(\Sigma_{L})$:
\begin{eqnarray*}
p_{a}&=&\gamma_{ab}(q)\,\dot q^{b}-p_{\alpha}\,\Gamma^{\alpha}_{a}(q)\; ,\\
p_i&=&\frac{1}{2}\frac{\partial \gamma_{ab}}{\partial q^i}\dot{q}^a\dot{q}^b-\frac{\partial V}{\partial q^i}-p_{\alpha}\frac{\partial \Gamma^{\alpha}_{a}}{\partial q^i}(q)\,\dot q^{a}\; ,\\
\dot{q}^{\alpha}&=&\Gamma^{\alpha}_{a}(q)\,\dot q^{a}\; .
\end{eqnarray*}
 The Legendre transformation is defined by
 $\F L=\tau_{T^{*}M}\circ(\alpha^{-1}_{M})\big|_{\Sigma_{L}}$,
 or locally by:
 \[
 \F L(q^i, \dot{q}^a, \tilde{\mu}_{\alpha})=(q^i, \gamma_{ab}\dot{q}^a-\tilde{\mu}_{\alpha}\,\Gamma^{\alpha}_{a}(q), \tilde{\mu}_{\alpha})\; .
 \]
Since $\left(\frac{\partial^2 L}{\partial \dot{q}^a\partial \dot{q}^b}-\tilde{\mu}_{\alpha}
\frac{\partial^2 \Psi^\alpha}{\partial \dot{q}^a\partial \dot{q}^b}\right)=(\gamma_{ab})
$ the constrained system $(L, C)$ is regular.

Moreover, the energy function $E_L: \Sigma_L\rightarrow \R$ is precisely
\[
E_L(q^i, \dot{q}^a, \tilde{\mu}_{\alpha})=\frac{1}{2}\gamma_{ab}\dot{q}^a\dot{q}^b-V(q)\; .
\]
Therefore, the Hamiltonian function can be expressed by $H=E_{L}\circ(\F L)^{-1}: T^*M\to \R$
\begin{equation}\label{Hamil}
H(q\,,\,p)=\frac{1}{2}\,\gamma^{ab}(q)\,P_{a}\,P_{b}+V(q),
\end{equation}
where $\gamma_{a\,b}\,\gamma^{b\,c}=\delta_{a}^{c}$ and $P_{a}=p_{a}+p_{\alpha}\,\Gamma^{\alpha}_{a}(q)$.

\subsubsection{Discretization: symplectic Euler method}

Taking into account equations (\ref{Lag}) and (\ref{Cons}), we define the discrete Lagrangian $\mathbb{L}_{d}:M\times M\Flder\R$ and the set of independent constraints in the following way (see \cite{MarsdenWest} for more details):
\begin{eqnarray}\nonumber
&&\mathbb{L}_{d}(q_{0},q_{1})=h\mathbb{L}(q_{0},\frac{q_{1}-q_{0}}{h})=
\frac{1}{2h}g_{i\,j}(q_0)(q^i_{1}-q^i_{0})( q^j_{1}-q^j_{0})-hV\lp q_{0}\rp,\\\label{EulerLag}\\
\nonumber
&&\lp\frac{q^{\alpha}_{1}-q^{\alpha}_{0}}{h}\rp=\Gamma^{\alpha}_{a}(q_{0})
\lp\frac{q^a_{1}-q^a_{0}}{h}\rp.
\end{eqnarray}
Next,  we can explicitly obtain the coordinates for the submanifold $\Upsilon(\Sigma_{L_{d}})$ given in equations (\ref{UpSigmaL}), namely
\begin{small}
\begin{eqnarray}\label{IntEuler}\nonumber
\lp p_{0}\rp_{a}&=&\frac{1}{h}g_{a\,j}(q_0)( q^j_{1}-q^j_{0})+h\der_{a}V\lp q_{0}\rp
-\frac{1}{2h}\partial_a g_{i\,j}(q_0)( q^i_{1}-q^i_{0})( q^j_{1}-q^j_{0})
\\
&&+\lp\lambda_{1}\rp_{\beta}\der_{a}\Gamma^{\beta}_{b}(q_{0})\lp q^b_{1}-q^b_{0}\rp-(\lambda_1)_{\alpha}\Gamma^{\alpha}_{a}(q_{0}),\\\nonumber
\\\nonumber
\lp p_{0}\rp_{\alpha}&=&\frac{1}{h}g_{\alpha\,j}(q_0)( q^j_{1}-q^j_{0})+h\der_{\alpha}V\lp q_{0}\rp-\frac{1}{2h}\partial_{\alpha}g_{i\,j}(q_0)( q^i_{1}-q^i_{0})( q^j_{1}-q^j_{0})
\\
&&+(\lambda_{1})_{\alpha}+(\lambda_{1})_{\beta}\der_{\alpha}\Gamma^{\beta}_{b}(q_{0})( q^b_{1}-q^b_{0}),\\\nonumber
\\
\lp p_{1}\rp_{a}&=&\frac{1}{h}g_{a\,j}(q_0)( q^j_{1}-q^j_{0})-(\lambda_{1})_{\alpha}\Gamma^{\alpha}_{a}(q_{0}),\\\nonumber
\\\nonumber
\lp p_{1}\rp_{\alpha}&=&\frac{1}{h}g_{\alpha\,j}(q_0)( q^j_{1}-q^j_{0})+(\lambda_{1})_{\alpha},\\\nonumber
\\\nonumber
\lp q^{\alpha}_{1}-q^{\alpha}_{0}\rp&=&\Gamma^{\alpha}_{a}( q_{0})( q^a_{1}-q^a_{0}),
\end{eqnarray}
\end{small}
where $\der_{a}$, $\der_{\alpha}$ mean $\frac{\der}{\der q^{a}}$ and $\frac{\der}{\der q^{\alpha}}$, respectively. It is important to note that (\ref{IntEuler}) is a set of $2m+n$ equations with $2m+n$ unknowns, which are $\lp q_{1}\rp^{a},\lp q_{1}\rp^{\alpha},\lp p_{1}\rp_{a}$, $\lp p_{1}\rp_{\alpha}$ and $\lp\lambda_{1}\rp_{\alpha}$.

Alternatively, we can apply the  so-called Euler symplectic method (see \cite{Hairer})
\begin{equation}\label{EulerA}
p_{1}=p_{0}-h\frac{\der H}{\der q}\lp q_{0}, p_{1}\rp,\,\,\,\,\,q_{1}=q_{0}+h\frac{\der H}{\der p}\lp q_{0},p_{1}\rp
\end{equation}
to the Hamiltonian function $H: T^{*}M\Flder\R$ defined in (\ref{Hamil}).
We deduce the  following set of equations:
\begin{small}
\begin{eqnarray}\nonumber
\lp q_{1}\rp^{a}&=&\lp q_{0}\rp^{a}+h\gamma^{ab} \lp P_{1}\rp_{b},\\\nonumber\\\nonumber
\lp q_{1}\rp^{\alpha}&=&\lp q_{0}\rp^{\alpha}+h\Gamma^{\alpha}_{a}\gamma^{ab}\lp P_{1}\rp_{b},\\\label{MetNum}\\\nonumber
\lp p_{1}\rp_{a}&=&\lp p_{0}\rp_{a}-h\der_{a}V-h\lc\frac{1}{2}\lp\der_{a}\gamma^{bc}\rp\lp P_{1}\rp_{b}\lp P_{1}\rp_{c}+\gamma^{bc}\lp P_{1}\rp_{b}\der_{a}\lp P_{1}\rp_{c}\rc,
\\\nonumber\\\nonumber
\lp p_{1}\rp_{\alpha}&=&\lp p_{0}\rp_{\alpha}-h\der_{\alpha}V-h\lc\frac{1}{2}\lp\der_{\alpha}\gamma^{ab}\rp\lp P_{1}\rp_{a}\lp P_{1}\rp_{b}+\gamma^{ab}\lp P_{1}\rp_{a}\der_{\alpha}\lp P_{1}\rp_{b}\rc,
\end{eqnarray}
\end{small}
where $\lp P_{1}\rp_{a}=\lp p_{1}\rp_{a}+\lp p_{1}\rp_{\alpha}\Gamma^{\alpha}_{a}$, and $V$, $\gamma^{ab}$, $\Gamma^{\alpha}_{a}$, $\lp P_{1}\rp_{a}$ are evaluated at $q_{0}$.
Regarding equations (\ref{IntEuler}), is easy to express $\lambda_{1}$ in terms of $p$ and $q$. Since
\begin{equation}\label{Truquele}
0=\der_{a}\delta_{d}^{c}=\der_{a}\lp\gamma^{cb}\gamma_{bd}\rp=\lp\der_{a}\gamma^{cb}\rp\gamma_{bd}+\gamma^{cb}\lp\der_{a}\gamma_{bd}\rp,
\end{equation}
and  $\lp\der_{a}\gamma^{cb}\rp\gamma_{bd}=-\gamma^{cb}\lp\der_{a}\gamma_{bd}\rp$, is easy to check, after a straightforward calculation, that equations (\ref{IntEuler}) reduce to (\ref{MetNum}).

\subsubsection{Discretization: Midpoint rule}
 Define the discrete Lagrangian and the discrete constraints using the {\it midpoint rule} (see \cite{MarsdenWest} for more details), that is:
\begin{small}
\begin{eqnarray}\nonumber
&&\mathbb{L}_{d}(q_{0},q_{1})=h\mathbb{L}(\qmid,\frac{q_{1}-q_{0}}{h})=\\\label{MidLag}
&&\frac{1}{2h}g_{i\,j}(\qmid)(q_{1}-q_{0})^{i}(q_{1}-q_{0})^{j}-hV(\qmid),\\\nonumber \\\nonumber
&&\lp\frac{q_{1}-q_{0}}{h}\rp^{\alpha}=\Gamma_{a}^{\alpha}(\qmid)\lp\frac{q_{1}-q_{0}}{h}\rp^{a}.
\end{eqnarray}
\end{small}
Now, we can explicitly obtain the corresponding equations derived from the submanifold $\Upsilon(\Sigma_{L_{d}})$.
A straightforward computation shows that these equations are equivalent to the corresponding ones derived from the so-called midpoint rule
\begin{small}
\begin{equation}\label{MidRule}
q_{1}=q_{0}+h\frac{\der H}{\der p}\lp\qmid\,,\,\pmid\rp,\,\,\,\,\,p_{1}=p_{0}-h\frac{\der H}{\der q}\lp\qmid\,,\,\pmid\rp,
\end{equation}
\end{small}
which is a symplectic method of order $2$.

\begin{equation}\label{SuperDiagrama}
\xymatrix{
\Sigma_{L}\ar[dd]_{\phi_{L}}\ar@{-->}[r]^{\varphi_{d}} &\Sigma_{L_{d}}\ar[dd]_{\phi_{L_{d}}}\ar[rrr]^{\F L_{d}^{-}}& &&
T^{*}M\ar[dd]<3pt>^{(\phi_{H})_{d}}\ar[dd]<-3pt>_{\phi_{_{\F L_{d}^{-}}}} & T^{*}M\ar[dd]^{\phi_{H}}\ar@{-->}[l]_{\hat\varphi_{d}}\\
 & & & & \\
\Sigma_{L}\ar@{-->}[r]_{\varphi_{d}} &\Sigma_{L_{d}}\ar[rrr]^{\F L_{d}^{-}}& && T^{*}M &T^ {*}M\ar@{-->}[l]^{\hat\varphi_{d}}
}
\end{equation}
\begin{equation}\label{SuperDiagrama2}
\xymatrix{
\Sigma_{L}\ar[dd]_{\phi_{L}}\ar@{-->}[r]^{\varphi_{d}} &\Sigma_{L_{d}}\ar[dd]_{\phi_{L_{d}}}\ar[rrr]^{\F L_{d}^{+}}& &&
T^{*}M\ar[dd]<3pt>^{(\phi_{H})_{d}}\ar[dd]<-3pt>_{\phi_{_{\F L_{d}^{+}}}} & T^{*}M\ar[dd]^{\phi_{H}}\ar@{-->}[l]_{\hat\varphi_{d}}\\
 & & & & \\
\Sigma_{L}\ar@{-->}[r]_{\varphi_{d}} &\Sigma_{L_{d}}\ar[rrr]^{\F L_{d}^{+}}& && T^{*}M &T^ {*}M\ar@{-->}[l]^{\hat\varphi_{d}}
}
\end{equation}

The results of the previous examples can be summarized in the diagrams (\ref{SuperDiagrama}) and \eqref{SuperDiagrama2}, which are explained in the following lines.
\begin{itemize}
\item {\bf From right to left}: $\phi_{H}$ is the Hamiltonian flow derived from (\ref{HamVecFi}) applied to a Hamiltonian function $H$, $\hat\varphi_{d}$ is the discretization of that flow (concretely the symplectic Euler method (\ref{EulerA}) and the midpoint rule (\ref{MidRule})), $(\phi_{H})_{d}$ is the discrete flow in $T^{*}M$ provided by $\hat\varphi_{d}$.

\item {\bf From left to right}: $\phi_{L}$ is the constrained Lagrangian flow resulting from equations (\ref{Vako}) applied to the continuous Lagrangian, $\varphi_{d}$ is the discretization applied to that Lagrangian ((\ref{EulerLag}) and (\ref{MidLag})), $\phi_{L_{d}}$ is the discrete flow within $\Sigma_{L_{d}}$ (\ref{SigmaLd}) due to $\varphi_{d}$, $\phi_{\F L_{d}^{\pm}}$ is the discrete flow in $T^{*}M\times T^{*}M$ (\ref{UpsSigmaLd}).
\end{itemize}
In order to be more explicit, $\hat\varphi_d:T^*M\Flder T^*M$ represents the discrete flow generated by applying a symplectic method to the Hamiltonian equations (e.g. \eqref{EulerA} and \eqref{MidRule}). On the other hand, $\varphi_d$ represents the discretization mapping, that is $\varphi_d:TM\Flder M\times M$, e.g. $\varphi_d(q,\dot q)=(q_k,\frac{q_{k+1}-q_k}{h})$ or $\varphi_d(q,\dot q)=(\frac{q_{k+1}+q_k}{2},\frac{q_{k+1}-q_k}{h})$.

As expected from Theorems \ref{TulczyAlpha} and \ref{Discreto}, what we explicitly show is that $\phi_{\F L_{d}^{\pm}}=(\phi_{H})_{d}$ using some discretizations (we have depicted the particular cases of the symplectic Euler methods and the midpoint rule). In other words, the diagrams (\ref{SuperDiagrama}) and \eqref{SuperDiagrama2} are commutative in those particular cases, as also is when the discretization $\hat\varphi_d$ corresponds to the exact discrete Lagrangian's (see \cite{MarsdenWest}). Therefore, using a discrete variational integrator for the constrained continuous Lagrangian system or applying  a symplectic integrator to the associated continuous Hamiltonian problem are equivalent approaches.



\subsection{The Martinet case: symplectic integrators for  sub-Riemannian geometry}
Let us consider the Hamiltonian function $H:T^*\R^{3}\Flder\R$
\begin{equation}\label{MartHam}
H(q,p)=\frac{1}{2}\lp\lp p_{x}+p_{z}\frac{y^{2}}{2}\rp^{2}+\frac{p_{y}^{2}}{(1+\beta\,x)^{2}}\rp,
\end{equation}
where $q=(x,y,z)^{T}\in\R^{3}$ and $p=(p_{x},p_{y},p_{z})\in(\R^{3})^{*}\simeq\R^{3}$. From (\ref{MartHam}) we can locally define $X_{H}(T^{*}M)$, particulary $X_{H}(T^*\R^{3})$, through the Hamiltonian equations, i.e:
\begin{equation}\label{XH}
\begin{array}{cccccccc}
\dot x&=&p_{x}+p_{z}\frac{y^{2}}{2},&&&\dot p_{x}&=&\frac{\beta\,p_{y}^{2}}{(1+\beta\,x)^{3}},\\
\dot y&=&\frac{p_{y}}{(1+\beta\,x)^{2}},&&&\dot p_{y}&=&-\lp p_{x}+p_{z}\frac{y^{2}}{2}\rp\,p_{z}\,y,\\
\dot z&=&\lp p_{x}+p_{z}\frac{y^{2}}{2}\rp\,\frac{y^{2}}{2},&&&\dot p_{z}&=&0.
\end{array}
\end{equation}
 The associated Legendre transform $\F H$ is in  this particular case  written as
\begin{small}
\begin{equation}\label{FH}
\F H(x,y,z;p_{x},p_{y},p_{z})=(x,y,z\,;\,(p_{x}+p_{z}\frac{y^{2}}{2}),\frac{p_{y}}{(1+\beta\,x)^{2}},\lp p_{x}+p_{z}\frac{y^{2}}{2}\rp\,\frac{y^{2}}{2}).
\end{equation}
\end{small}
Looking at (\ref{XH}) and (\ref{FH}) is easy to realize that
\[
C\subset T\R^{3}=\lc(x,y,z\,;\,\dot x,\dot y,\dot z)\,\,\,{\mbox{s.t.}}\,\,\,\,\dot z=\frac{y^{2}}{2}\,\dot x\rc.
\]
Next, we will obtain the  Lagrangian function $L: C\rightarrow \R$ using the implicit equation given in  (\ref{lagrangian}):
\[
L\circ\F H=\frac{1}{2}\lp\lp p_{x}+p_{z}\frac{y^{2}}{2}\rp^{2}+\frac{p_{y}^{2}}{(1+\beta\,x)^{2}}\rp.
\]
Finally, using $\F H$ we arrive to
\begin{equation}\label{MartLag}
L(x, y, z, \dot{x}, \dot{y})=\frac{1}{2}\lp\dot x^{2}+(1+\beta\,x)^{2}\,\dot y^{2}\rp.
\end{equation}
Consequently, our approach allows us to conclude that the Hamiltonian system (\ref{MartHam}) is equivalent to the Lagrangian one (\ref{MartLag}) subject to the constraints $\dot z=\frac{y^{2}}{2}\,\dot x$. We clearly recognize in (\ref{MartLag}) a Martinet sub-Riemannian structure (\cite{Agra}, \cite{Bonn}, \cite{Vilmart}), which is described by the triple $(U\,,\,\Delta\,,\,g)$. In this triple, $U$ is an open neighborhood of the origin in $\R^{3}$, $\Delta$ is a distribution corresponding to $\Delta=\mbox{ker}\,\alpha$ for $\alpha=\mbox{d}z-\frac{y^{2}}{2}\mbox{d}x$  and $g$ is a Riemannian metric.
In  the particular case $g=\mbox{d}x^{2}+(1+\beta\,x)^{2}\mbox{d}y^{2}$, we deduce that  (\ref{MartLag}) corresponds to $L={\mathbb L}\big|_C$ where ${\mathbb L}(q,\dot q)=\frac{1}{2}\,g\,(\frac{\der}{\der q}\,,\,\frac{\der}{\der q})$, where $\frac{\der}{\der q}=\dot x\frac{\der}{\der x}+\dot y\frac{\der}{\der y}+\dot z\frac{\der}{\der z}$. In addition,  the constraints are given by $\alpha\,(\frac{\der}{\der q})=0$.

\subsubsection{Discrete Case}

Let us consider the symplectic Euler method (\ref{EulerA}). Is easy to see that  a type-2 generating function of the  approximated Hamiltonian  flow is
\begin{equation}\label{Hmas}
H^{+}(q_{0},p_{1})=q_{0}\,p_{1}+h\,H(q_{0},p_{1}),
\end{equation}
where $H$ is the Hamiltonian function $H:T^{*}\R^3\Flder\R$. In other words:
\[
p_{0}=\frac{\der\,H^{+}(q_{0},p_{1})}{\der q_{0}},\,\,\,\,\,\,\,q_{1}=\frac{\der\,H^{+}(q_{0},p_{1})}{\der p_{1}}.
\]
Under these considerations, we can define the local coordinates for $\Ups$:
\[
\Ups=\lc q_{0}\,\,,\,\,\frac{\der\,H^{+}(q_{0},p_{1})}{\der q_{0}}\,\,,\,\,\frac{\der\,H^{+}(q_{0},p_{1})}{\der p_{1}}\,\,,\,\,p_{1}\rc.
\]
Now,  projecting $\Ups$  onto $\R^3\times \R^3$, we obtain that
\begin{eqnarray}\nonumber
x_{1}&=&x_{0}+h\lp(p_{1})_{x}+(p_{1})_{z}\frac{y_{0}^{2}}{2}\rp,\\\label{Projection}
y_{1}&=&y_{0}+h\frac{(p_{1})_{y}}{(1+\beta\,x_{0})^{2}},\\\nonumber
z_{1}&=&z_{0}+h\lp(p_{1})_{x}+(p_{1})_{z}\frac{y_{0}^{2}}{2}\rp\frac{y_{0}^{2}}{2}.
\end{eqnarray}
From the last equations we obtain the constraint $(z_{1}-z_{0})=\frac{y^{2}_{0}}{2}(x_{1}-x_{0})$, which defines the submanifold
\[
C_{d}=\{ (x_0, y_0, z_0; x_1, y_1, z_1)\in \R^3\times \R^3\; |\; (z_{1}-z_{0})=\frac{y^{2}_{0}}{2}(x_{1}-x_{0})\}.
\]

The next step to completely determine the discrete constrained Lagrangian system is to obtain $L_{d}$. In that sense, we take the  usual generating function $S$ mentioned in Theorem \ref{Discreto}.
Transforming $S$ into a {\it  type-2 generating function} (\cite{Aitor}) we arrive to the implicitly defined expression: $S(q_{0},p_1)=p_{1}\,q_{1}-H^{+}(q_{0},p_{1})$, which, taking into account (\ref{Hmas}) leads to
\begin{equation}\label{as}
S(q_{0},p_{1})=h\lp p_{1}\frac{\der H(q_{0},p_{1})}{\der p_{1}}-H(q_{0},p_{1})\rp.
\end{equation}
We have shown in Theorem \ref{Discreto} that $S$ is $\pi_{M\times M}\big|_{\Ups}$-projectable onto $L_{d}:C_{d}\Flder\R$. Thus, from (\ref{as}) and according to equations (\ref{Projection}) we finally arrive to
\begin{equation}\label{MartLagDisc}
L_{d}(q_{0},q_{1})=\frac{1}{2h}\lp(x_{1}-x_{0})^{2}+\frac{y_{0}^{2}}{2}(y_{1}-y_{0})^{2}\rp,
\end{equation}
for
\begin{equation}\label{SubDisc}
C_{d}\subset M\times M=\lc (q_{0}\,,\,q_{1})\,\,|\,\,(z_{1}-z_{0})=\frac{y^{2}_{0}}{2}(x_{1}-x_{0})\rc.
\end{equation}

This is the expected result as is easily seen taking into account the continuous Lagrangian (\ref{MartLag}) and the submanifold $C\subset TM$ defined by the continuous constraint $\dot z=\frac{y^{2}}{2}\,\dot x$. If we define both the discrete Lagrangian and the discrete submanifold $C_{d}$ in the usual symplectic Euler discretization (\cite{MarsdenWest}), i.e. $L_{d}(q_{0},q_{1})=h\,L(q_{0},\frac{q_{1}-q_{0}}{h})$, we easily obtain (\ref{MartLagDisc}) and
(\ref{SubDisc}).

\subsection{Symplectic St\"ormer-Verlet method}\label{SVer}

Due to its well-behaved features, namely reversibility, symplecticity, volume preservation and conservation of first integrals, the St\"ormer-Verlet method is one of the most important examples in geometric numerical integration (see \cite{Hairer} and references therein).
For a Hamiltonian system determined by $H: T^*M\rightarrow \R$, the St\"ormer-Verlet method reads
\begin{eqnarray}\nonumber
\pmed&=&p_{k}-\frac{h}{2}H_{q}(q_{k},\pmed),\\\label{Sverlet}
q_{k+1}&=&q_{k}+\frac{h}{2}\lp H_{p}(q_{k},\pmed)+H_{p}(q_{k+1},\pmed)\rp,\\\nonumber
p_{k+1}&=&\pmed-\frac{h}{2}H_{q}(q_{k+1},\pmed),
\end{eqnarray}
where $q_{k}\in\R^{n}$, $p_{k}\in\lp\R^{n}\rp^{*}$ and $H_{q}, H_p$ are the derivatives of the Hamiltonian function respect $q$ and $p$, respectively. As we did for the momenta, we can fix an intermediate configuration point $\qmed=q_{k}+\frac{h}{2}H_{p}(q_{k},\pmed)$ and consider (\ref{Sverlet}) as a two step integrator:
\begin{eqnarray}\nonumber
\qmed&=&q_{k}+\frac{h}{2}H_{p}(q_{k},\pmed),\\ \vspace{-1.3cm}\label{Mas}\\\nonumber
\pmed&=&\pk-\frac{h}{2}H_{q}(\qk,\pmed),\\\nonumber\\\nonumber\\\nonumber
q_{k+1}&=&\qmed+\frac{h}{2}H_{p}(q_{k+1},\pmed),\\\vspace{-1.3cm}\label{Menos}\\\nonumber
p_{k+1}&=&\pmed-\frac{h}{2}H_{q}(q_{k+1},\pmed).
\end{eqnarray}

Equations (\ref{Mas}) and (\ref{Menos}) show the well-known fact that the St\"ormer-Verlet method is the composition of two different symplectic Euler schemes. In addition, it is clear that they are respectively generated by the 2- and 3-type generating functions
\begin{eqnarray*}
H^{+}(\qk,\pmed)&=&\pmed\,\qk+\frac{h}{2}H(\qk,\pmed),\\
H^{-}(q_{k+1},\pmed)&=&\pmed\,q_{k+1}-\frac{h}{2}H(q_{k+1},\pmed),
\end{eqnarray*}
which, taking into account that $p_{1}\,\mathrm{d}q_{1}-p_{0}\,\mathrm{d}q_{0}=\mathrm{d} S(q_{0},q_{1})$, lead to
\begin{small}
\begin{eqnarray}\label{Smas}
S^{+}(\qk,\pmed)&=&\pmed\,\qmed-\pmed\,\qk-\frac{h}{2}H(\qk,\pmed),\\\label{Smenos}
S^{-}(\pmed,q_{k+1})&=&-\pmed\,\qmed+\pmed\,q_{k+1}
-\frac{h}{2}H(q_{k+1},\pmed).
\end{eqnarray}
\end{small}
Now, as shown in \cite{Aitor}, we can construct a 1-type generating function $S(\qk,q_{k+1})$ by
\begin{eqnarray}\label{ESE}
&&S(\qk,q_{k+1})=S^{+}(\qk,\pmed)+S^{-}(\pmed,q_{k+1})\\\nonumber
&&\ =\pmed\,q_{k+1}-\pmed\,\qk-\frac{h}{2}\lp H(\qk,\pmed)+H(q_{k+1},\pmed)\rp,
\end{eqnarray}
and a extremal condition in the intermediate variable $\pmed$. That is,
\begin{eqnarray*}
&&\mathrm{d} S=\frac{\der S}{\der\qk}\de\qk+ \frac{\der S}{\der\qkm}\de\qkm\\
&&=\lc\frac{\der S^{+}}{\der\qk}+\lp\frac{\der S^{+}}{\der\pmed}+\frac{\der S^{-}}{\der\pmed}\rp\frac{\der\pmed}{\der\qk}\rc\de\qk+\\
&&\lc\frac{\der S^{-}}{\der\qkm}+\lp\frac{\der S^{+}}{\der\pmed}+\frac{\der S^{-}}{\der\pmed}\rp\frac{\der\pmed}{\der\qkm}\rc\de\qkm,
\end{eqnarray*}
which leads to
\begin{equation}\label{Condicion}
\frac{\der S^{+}}{\der\pmed}+\frac{\der S^{-}}{\der\pmed}=0,
\end{equation}
(put in another way, $S(q_{k},q_{k+1})$ is not a function of $\pmed$ and consequently its partial derivative with respect to this variable should vanish). In other words, we obtain the first equation in (\ref{Sverlet}) by $-p_{k}=\frac{\der S(\qk,q_{k+1})}{\der\qk}$, the third one by $p_{k+1}=\frac{\der S(\qk,q_{k+1})}{\der q_{k+1}}$, and the second one by $\frac{\der S(\qk,q_{k+1})}{\der \pmed}=0$.

As shown in Theorem \ref{Discreto}, $S(\qk,\qkm)$ is projectable onto $L_{d}$, while condition (\ref{Condicion}) provides $C_{d}\subset M\times M$.

\subsubsection{Regular systems} Consider the usual mechanical Hamiltonian
\[
H(q,p)=\frac{1}{2}\,p\,M^{-1}\,p^{T}+V(q),
\]
where $M$ is a symmetric regular $n\times n$ matrix. From the second equation in (\ref{Sverlet}) is easy to check that $\pmed^{T}=M(\frac{q_{k+1}-q_{k}}{h})$. Hence, projecting (\ref{ESE}) onto $M\times M$ we arrive to
\[
L_{d}(\qk,\qkm)=\frac{1}{2h}(\frac{q_{k+1}-q_{k}}{h})^{T}\,M\,(\frac{q_{k+1}-q_{k}}{h})-\frac{h}{2}\lp V(\qk)+V(\qkm)\rp.
\]
In the expression just above, we clearly recognize the discretization
\[
L_{d}(\qk,\qkm)=\frac{h}{2}L(q_{k},\frac{\qkm-\qk}{h})+\frac{h}{2}L(\qkm,\frac{\qkm-\qk}{h})
\]
for the usual mechanical Lagrangian $L(q,\dot q)=\frac{1}{2}\,\dot q^{T}\,M\,\dot q-V(q)$.

\subsubsection{Martinet structure} Consider again the sub-Riemannian Martinet structure in (\ref{MartHam}). Recall that $q=(x,y,z)^T\in\R^{3}$ and $p=(p_x,p_y,p_z)\in \R^3$. From (\ref{Condicion}) we obtain
\begin{footnotesize}
\begin{eqnarray*}
x_{k+1}&=&x_{k}+\frac{h}{2}\lc\lp(\pmed)_{x}+(\pmed)_{z}\frac{y_{k}^{2}}{2}\rp+\lp(\pmed)_{x}+(\pmed)_{z}\frac{y_{k+1}^{2}}{2}\rp\rc,\\\\
y_{k+1}&=&y_{k}+\frac{h}{2}\lc\frac{(\pmed)_{y}}{(1+\beta\,x_{k})^{2}}+\frac{(\pmed)_{y}}{(1+\beta\,x_{k+1})^{2}}\rc,\\\\
z_{k+1}&=&z_{k}+\frac{h}{2}\lc\lp(\pmed)_{x}+(\pmed)_{z}\frac{y_{k}^{2}}{2}\rp\frac{y_{k}^{2}}{2}+\lp(\pmed)_{x}+(\pmed)_{z}\frac{y_{k+1}^{2}}{2}\rp\frac{y_{k+1}^{2}}{2}\rc.
\end{eqnarray*}
\end{footnotesize}


 As above, we consider a {\it two steps} (of $h/2$ size) interpretation of this sub-Riemannian system through equations (\ref{Mas}) and (\ref{Menos}) in the following manner (we denote $\pmed$ by $p$ for sake of simplicity):


\[
\begin{array}{c|c}

(\qk,\pk)\Flder(\qmed,\pmed) & (\qmed,\pmed)\Flder(\qkm,\pkm)\\ \\ \\

x_{k+1/2}=x_{k}+\frac{h}{2}\lp p_{x}+ p_{z}\frac{y_{k}^{2}}{2}\rp &   x_{k+1}=x_{k+1/2}+\frac{h}{2}\lp p_{x}+ p_{z}\frac{y_{k+1}^{2}}{2}\rp\\ \\

y_{k+1/2}=y_{k}+h\lc\frac{p_{y}}{(1+\beta\,x_{k})^{2}}\rc & y_{k+1}=y_{k+1/2}+h\lc\frac{p_{y}}{(1+\beta\,x_{k+1})^{2}}\rc \\ \\

z_{k+1/2}=z_{k}+\frac{h}{2}\lp p_{x}+p_{z}\frac{y_{k}^{2}}{2}\rp\frac{y_{k}^{2}}{2} &  z_{k+1}=z_{k+1/2}+\frac{h}{2}\lp p_{x}+p_{z}\frac{y_{k+1}^{2}}{2}\rp\frac{y_{k+1}^{2}}{2}\\ \\
\Downarrow &  \Downarrow\\ \\
C_d: \lp\frac{z_{k+1/2}-z_{k}}{h/2}\rp=\frac{y^2_k}{2}\lp\frac{x_{k+1/2}-x_{k}}{h/2}\rp  & C_d: \lp\frac{z_{k+1}-z_{k+1/2}}{h/2}\rp=\frac{y^2_{k+1}}{2}\lp\frac{x_{k+1}-x_{k+1/2}}{h/2}\rp\\\\
L^{+}_{d}(\qk,\qmed) &  L^{-}_{d}(\qmed,\qkm)
\end{array}
\]
\vspace{0.2cm}

As shown in the table above, both discrete submanifolds are respectively defined by the constraints
$\lp\frac{z_{k+1/2}-z_{k}}{h/2}\rp=\frac{y^2_k}{2}\lp\frac{x_{k+1/2}-x_{k}}{h/2}\rp$ and $\lp\frac{z_{k+1}-z_{k+1/2}}{h/2}\rp=\frac{y^2_{k+1}}{2}\lp\frac{x_{k+1}-x_{k+1/2}}{h/2}\rp$. Taking into account equations (\ref{Smas}) and (\ref{Smenos}) in $\S$ \ref{SVer}, define the two discrete Lagrangian functions (implicitly expressed) as
\[
L^{+}_{d}(\qk,\qmed)=\pmed\,\qmed-\pmed\,\qk-\frac{h}{2}H(\qk,\pmed),
\]
with  $\qmed=q_{k}+\frac{h}{2}H_{p}(q_{k},\pmed)$ (recall that this expression corresponds to the generating function after projecting onto $M\times M$). On the other hand
\[
L^{-}_{d}(\qmed,\qkm)=-\pmed\,\qmed+\pmed\,q_{k+1}
-\frac{h}{2}H(q_{k+1},\pmed),
\]
with   $q_{k+1}=\qmed+\frac{h}{2}H_{p}(q_{k+1},\pmed)$.

From the previous expressions and following the discussion in \cite{MarsdenWest} $\S$ 2.5.1, we divide each step $(\qk,\qkm)$ into 2 substeps $(\qk=\qk^0\,,\,\qk^1=\qmed)$ and $(\qk^1=\qmed\,,\, q_{k+1}^0=\qkm)$. Take the discrete action sum
\begin{eqnarray*}
\mathfrak{S}_{d}( \{\qk^0,\qk^1\}_{0}^{N-1})&=&\sum_{k=0}^{N-1}\lp L^{+}_{d}(\qk^{0},\qk^{1})+L^{-}_{d}(\qk^{1},q_{k+1}^{0})\rp\\&=&\sum_{k=0}^{N}\lp L^{+}_{d}(\qk,\qmed)+L^{-}_{d}(\qmed,\qkm)\rp.
\end{eqnarray*}
The corresponding Euler-Lagrange equations, resulting from requiring this action to be stationary, pair both neighbouring discrete Lagrangians together to give
\begin{eqnarray*}
D_2\,L_{d}^{+}(\qk^0.\qk^1)&+&D_1\,L_d^-(\qk^1,\qk^2)=0,\\
D_2\,L_{d}^{-}(\qk^1,\qk^2)&+&D_{1}\,L_{d}^{+}(\qkm^0,\qkm^{1})=0.
\end{eqnarray*}
 The equations just above completely determine the discrete dynamics in equations (\ref{Mas}) and (\ref{Menos}) for the Martinet sub-Riemannian system (\ref{MartHam}). In other words, the discrete scheme in (\ref{Mas}) and (\ref{Menos}) could be understood as two discrete Hamiltonian flows $F_{L_d^+}$ and $F_{L_d^-}$, each one of half-step $h/2$, respectively generated by the generating functions $L_d^+$ and $L_d^-$. The map over the entire time-step $h$ is thus the composition of the maps $F_{L_d^-}\circ F_{L_d^+}$.

\section{Conclusions and future work}

In this paper, we have carefully studied the relationship between Hamiltonian dynamics and constrained variational calculus, showing that under natural regularity conditions both are equivalent. The analysis is easily extended  to the discrete case, allowing us to explore the applications to the construction of symplectic integrators. Moreover, we have analyzed in parallel the case of classical nonholonomic mechanics in the discrete and continuous cases. Our technique allows us to extend the  comparison algorithms between nonholonomic and constrained variational cases to the discrete picture, finding additionally a new and simpler point of view in the continuous case.

Some interesting lines that we want to explore in future research are the following ones. First, many higher-order symplectic methods are obtained using composition of methods of lower-order. In our approach, this notion seems to be related with the notion of composability of canonical relations (which may fail even to be a manifold without appropriate transversality conditions \cite{Wei}). This study is a promising line for numerical simulation of Hamiltonian dynamics and also constrained systems.

One of the main problems of discrete  nonholonomic mechanics is the lack of preservation of geometric structures (non-preservation of the nonholonomic bracket, non-preservation of the nonholonomic momentum in general) mimicking the non-preservation of the continuous nonholonomic system.
Then, it is difficult to compare from this geometric perspective the nonholonomic integrators obtained from discrete nonholonomic mechanics with standard methods. One possibility is to use our comparison algorithms detecting if one particular nonholonomic integrator is preserving the common solutions for the continuous nonholonomic problem and its associated constrained variational problem.

Other interesting case that we want to explore is the  extension of our theory to reduced systems using the geometric framework given by the Lie algebroid and Lie groupoid formalisms \cite{Ma}. For instance, in the discrete case, our intention is to derive the dynamics using also Lagrangian submanifolds of the so-called tangent groupoid \cite{CDW}.
In this context we will start with a submanifold $N$ of a Lie groupoid $G\rightrightarrows M$, and a discrete Lagrangian $L_d: N\rightarrow \R$. Following Theorem \ref{tulchi}, we introduce the Lagrangian submanifold $\Sigma_{L_d}$ of the cotangent groupoid $T^*G$. The Hamiltonian side is determined by a Poisson flow defined on $A^*G$, the dual bundle of the  associated Lie algebroid $AG$ to $G$. Therefore, we have the following scheme:
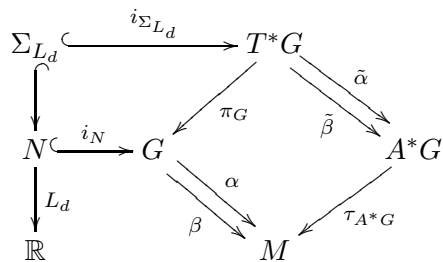
\[
\xymatrix{
\Sigma_{L_{d}}\ar@{^{(}->}[rr]^{i_{\Sigma_{L_{d}}}}\ar@{^{(}->}[d]&  &T^{*}G\ar[dl]^{\pi_{G}}\ar[dr]<3pt>^{\tilde\alpha}\ar[dr]<-3pt>_{\tilde\beta}&\\
N\ar@{^{(}->}[r]^{i_{N}}\ar[d]^{L_{d}} &G\ar[dr]<3pt>^{\alpha}\ar[dr]<-3pt>_{\beta} & &A^{*}G\ar[dl]^{\tau_{A^{*}G}}\\
\R& & M&
}
\]
which, in principle, will allow us to extend the theory presented in this paper.

\section*{Acknowledgements}

This work has been partially supported by MICINN (Spain) MTM2010-21186-C02-01 and  MTM2009-08166-E, project ``Ingenio
Mathematica'' (i-MATH) No. CSD 2006-00032 (Consolider-Ingenio 2010) and the European project IRSES-project ``Geomech-246981''. The authors are indebted to the referees of the former version of this paper for their comments, remarks and suggestions, which definitely have supposed a decisive improvement on our work.

\end{document}